\documentclass{natureprintstyle}
\usepackage{epsfig,caption}
\usepackage{color}
\usepackage{bm}
\usepackage{graphicx}
\usepackage{longtable}
\usepackage{amsmath}
\usepackage{amssymb}
\usepackage{rotating,xcolor}
\usepackage{hyperref}
\usepackage[switch]{lineno}
\hypersetup{
    hidelinks,
    colorlinks=false,
    urlcolor=black
}
\usepackage{multibib}
\newcites{meth}{Methods References}

\usepackage{color}
\usepackage{soul}
\usepackage{xspace}
\usepackage{hyperref}

\newcommand{\code}[1]{\texttt{#1}\xspace}

\newcommand{\unit}[1]{\ensuremath{\mathrm{\,#1}}\xspace}
\newcommand{\feh}{\unit{[Fe/H]}}

\newcommand{\logzzsun}{\ensuremath{\log Z/Z_\odot}\xspace}
\newcommand{\teff}{\ensuremath{T_\mathrm{eff}}\xspace}
\newcommand{\Teff}{\teff}
\newcommand{\logg}{\ensuremath{\log\,g}\xspace}
\newcommand{\alphafe}{\unit{[\alpha/Fe]}}

\newcommand{\kms}{\unit{km\,s^{-1}}}

\newcommand{\umpstarsdss}{SDSS~J0715$-$7334\xspace}
\newcommand{\umpstar}{J0715$-$7334\xspace}
\newcommand{\umpstartmass}{2MASS~J07153858$-$7334530\xspace}
\newcommand{\umpstargaia}{Gaia~DR3~5262850721755411072\xspace}
\newcommand{\umpstarsdssid}{SDSS\_ID~95803549\xspace}
\newcommand{\cdstar}{CD$-$38~245\xspace}

\newcommand{\caffaustar}{J1029$+$1729\xspace}
\newcommand{\kellerstar}{SMSS J0313$-$6708\xspace}

\title{A nearly pristine star from the Large Magellanic Cloud}

\author{Alexander~P.~Ji\textsuperscript{1, 2, 3, $\dagger$}, % 0000-0002-4863-8842
Vedant~Chandra\textsuperscript{4},                           % 0000-0002-0572-8012
Selenna~Mejias-Torres\textsuperscript{1},                    % 0009-0004-0052-6683
Zhongyuan~Zhang\textsuperscript{1},                          % 0009-0007-5954-7915
Philipp~Eitner\textsuperscript{5,6},                         % 0009-0007-4502-8081
Kevin~C.~Schlaufman\textsuperscript{7},                      % 0000-0001-5761-6779
Hillary~Diane~Andales\textsuperscript{1},                    % 0000-0002-2962-1391
Ha~Do\textsuperscript{1},                                    % 0009-0009-3497-8369
Natalie~M.~Orrantia\textsuperscript{1},                      % 0009-0006-4190-4964
Rithika~Tudmilla\textsuperscript{1},                         % 0009-0005-1505-0523
Pierre~N.~Thibodeaux\textsuperscript{1},                     % 0000-0002-3867-3927
Keivan~G.~Stassun\textsuperscript{8},                        % 0000-0002-3481-9052
Madeline~Howell\textsuperscript{9},                          % 0000-0003-0929-6541
Jamie~Tayar\textsuperscript{10},                             % 0000-0002-4818-7885
Maria~Bergemann\textsuperscript{5},                          % 0000-0002-9908-5571
Andrew~R.~Casey\textsuperscript{11,12},                      % 0000-0003-0174-0564
Jennifer~A.~Johnson\textsuperscript{9},                      % 0000-0001-7258-1834 
Joleen~K.~Carlberg\textsuperscript{13},                      % 0000-0001-5926-4471
William~Cerny\textsuperscript{14},                           % 0000-0003-1697-7062
Jos\'e~G.~Fern\'andez-Trincado\textsuperscript{15,16},          % 0000-0002-0900-9760
Keith~Hawkins\textsuperscript{17},                           % 0000-0002-1423-2174
Juna~A.~Kollmeier\textsuperscript{18},                       % 0000-0001-9852-1610
Chervin~F.~P.~Laporte\textsuperscript{19--22},               % 0000-0003-3922-7336
Guilherme~Limberg\textsuperscript{2,1},                        % 0000-0002-9269-8287
Tadafumi~Matsuno\textsuperscript{6},                         % 0000-0002-8077-4617
Szabolcs~M{\'e}sz{\'a}ros\textsuperscript{23--25},           % 0000-0001-8237-5209
Sean~Morrison\textsuperscript{26},                           % 0000-0002-6770-2627
David~L.~Nidever\textsuperscript{27},                        % 0000-0002-1793-3689
Guy~S.~Stringfellow\textsuperscript{28},                     % 0000-0003-1479-3059
Donald~P.~Schneider\textsuperscript{29,30},                  % 0000-0001-7240-7449
Riley~Thai\textsuperscript{11}                               % 0009-0000-9368-0006
}

\begin{document}
% \linenumbers
\maketitle
\let\thefootnote\relax\footnote{

\begin{affiliations}
\item Department of Astronomy \& Astrophysics, University of Chicago, 5640 S. Ellis Avenue, Chicago, IL 60637, USA
\item Kavli Institute for Cosmological Physics, University of Chicago, 5640 S Ellis Avenue, Chicago, IL 60637, USA
\item NSF-Simons AI Institute for the Sky (SkAI), 172 E. Chestnut St., Chicago, IL 60611, USA
$\dagger$ \href{mailto:alexji@uchicago.edu}{alexji@uchicago.edu}.
All author affiliations listed at end of paper.
\end{affiliations}
}

\vspace{-3.5mm}
\begin{abstract}
The first stars formed out of pristine gas, causing them to be so massive that none are expected to have survived until today. If their direct descendants were sufficiently low-mass stars, such stars could exist today and would be recognizable by having the lowest metallicities (abundance of elements heavier than helium). We present the independent identification and detailed chemical analysis of the star SDSS~\umpstar, finding ultra-low elemental abundances of both iron and carbon ($\mbox{[Fe/H]} = -4.3$, $\mbox{[C/Fe]} < -0.2$) and total metallicity $Z < 7.8 \times 10^{-7}$ ($\logzzsun < -4.3$). The star's orbit indicates that it originates from the halo of the Large Magellanic Cloud. Its heavy element abundance pattern can be explained by a primordial supernova with an initial mass of 30 solar masses. This star is over ten times more chemically pristine than the most extreme high-redshift galaxies currently found by the James Webb Space Telescope. It is sufficiently metal-poor that current models of low-mass star formation require dust cooling to explain its existence.
\end{abstract}

%%%%%%%%%%%%%%%%%%%%%%%%%%%%%%%%%%%%%%%%%%%%%%%%%%%%%

The first metal-free (Population~III) stars are thought to be unusually massive and thus all died in the early universe\cite{Bromm2009,Klessen2023}. 
Their supernovae synthesized the first metals, fundamentally changing the subsequent thermal evolution of gas and promoting gas fragmentation\cite{Bromm2003, Schneider2003}. 
Thus, the direct descendants of the first stars could have been sufficiently low mass ($\lesssim 0.8 M_\odot$) to survive to the present day and be found in the Milky Way galaxy\cite{Frebel2015}. 
The lowest metallicity star previously known is \caffaustar\cite{Caffau2011}, a star in the thick disk of the Milky Way with $Z < 1.66 \times 10^{-6}$ ($\logzzsun < -4.0$)\cite{Caffau2024}. 
While other stars with lower iron abundances have been found\cite{Christlieb2002,Frebel2005,Keller2014}, they have high carbon abundances and thus higher total metallicities ($\logzzsun > -3$). 
A few candidates have been suggested to be as metal-poor as {\caffaustar}\cite{Starkenburg2018,Limberg2025}, but without sufficiently stringent carbon abundance constraints to be certain.

\begin{figure}[h!]
\centering
\includegraphics[width=\linewidth]{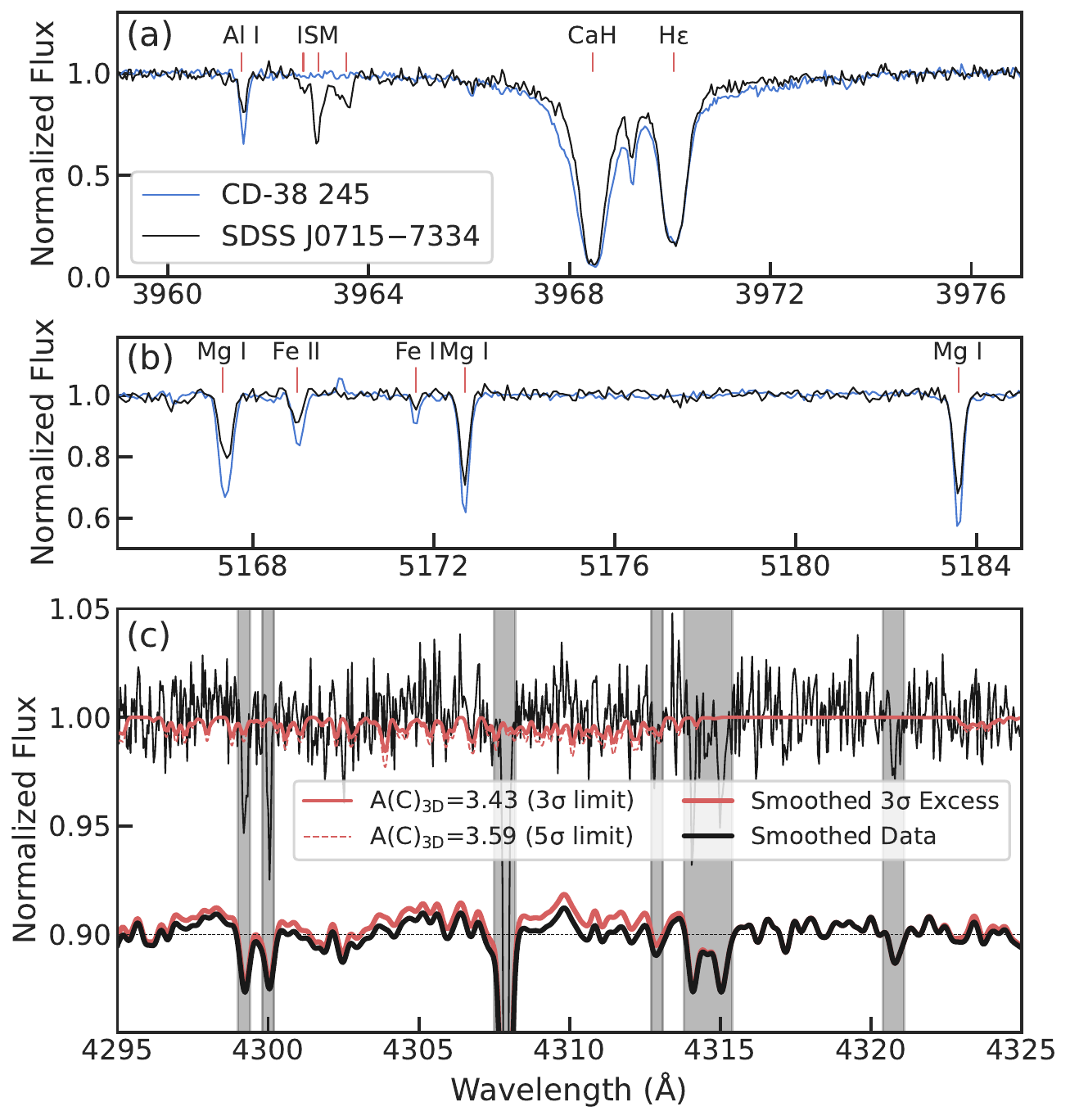}
\vspace{-4mm}
\caption{\textbf{\umpstar MIKE Spectrum.}
\textbf{(a)} Ca H and H$\varepsilon$ compared to {\cdstar} that has $\Teff=4889$\,K and $\feh=-3.9$ \cite{Roederer2014,Mittal2025} (blue line). \umpstar (black line) is narrower in both, indicating it is both cooler and more metal-poor. Interstellar medium (ISM) lines are also marked.
\textbf{(b)} Mg b region compared to \cdstar.
\textbf{(c)} CH G band region. Normal solid and thin dashed red lines indicate synthetic 3D LTE spectra corresponding to a $3\sigma$ and $5\sigma$ upper limit, respectively. Grey regions are masked as the 3D model only includes CH lines. The thick solid lines offset at $+0.9$ indicate the smoothed residuals for the $3\sigma$ upper limit (red) and the smoothed data assuming no carbon (black). The statistical significance of the upper limit is calculated using a profile likelihood and is related to the integrated area between these two lines (see Methods and \ref{fig:chsynth}).
\label{fig:spectrum}
}
\vspace{-6mm}
\end{figure}

SDSS~\umpstar was identified in the SDSS-V\cite{Kollmeier2026} Milky Way Mapper Halo survey using a \code{MINESweeper}\cite{Cargile2020} analysis of low-resolution BOSS spectra\cite{Chandra2025}. High-resolution followup was conducted using Magellan/MIKE\cite{Bernstein2003} (Figure~\ref{fig:spectrum}).
Figure~1c shows that no carbon G band is detected, an extraordinary finding for a cool red giant star ($\Teff = 4700 \pm 100\,$K, $\logg = 1.1 \pm 0.25$, $\feh_{\rm NLTE} = -4.3$; see Methods for details).
We conducted a chemical abundance analysis in 1D assuming Local Thermodynamic Equilibrium (LTE) and adopting photometric stellar parameters, then corrected the abundances of most elements for non-LTE (NLTE) effects\cite{Gerber2023}. 
We paid special attention to the carbon upper limit, using 3D LTE models of the CH molecule\cite{Eitner2024} and obtaining a $3\sigma$ upper limit $A(C)_{\rm 3D} < 3.43$.
Additionally, stellar interiors convert carbon into nitrogen, which is brought to the surface as stars ascend the red giant branch;
correcting for this effect with evolutionary models\cite{Placco2014} results in a final $A(C) < 3.99$ or $\mbox{[C/H]} < -4.52$.
The adopted chemical abundances including NLTE, 3D, and evolutionary corrections are presented in Table~\ref{tab:abunds}.
The lithium abundance is low, but this is due to depletion on the red giant branch rather than an indication of primordial lithium\cite{Lind2009}.
This same star was also identified by Limberg et al.\cite{Limberg2025} using \textit{Gaia} spectra.
Their 1D LTE abundance analysis of a lower quality spectrum is consistent with our analysis, but without a stringent carbon (and thus total metallicity) constraint (see Methods).

\begin{figure}[h!]
\centering
\includegraphics[width=\linewidth]{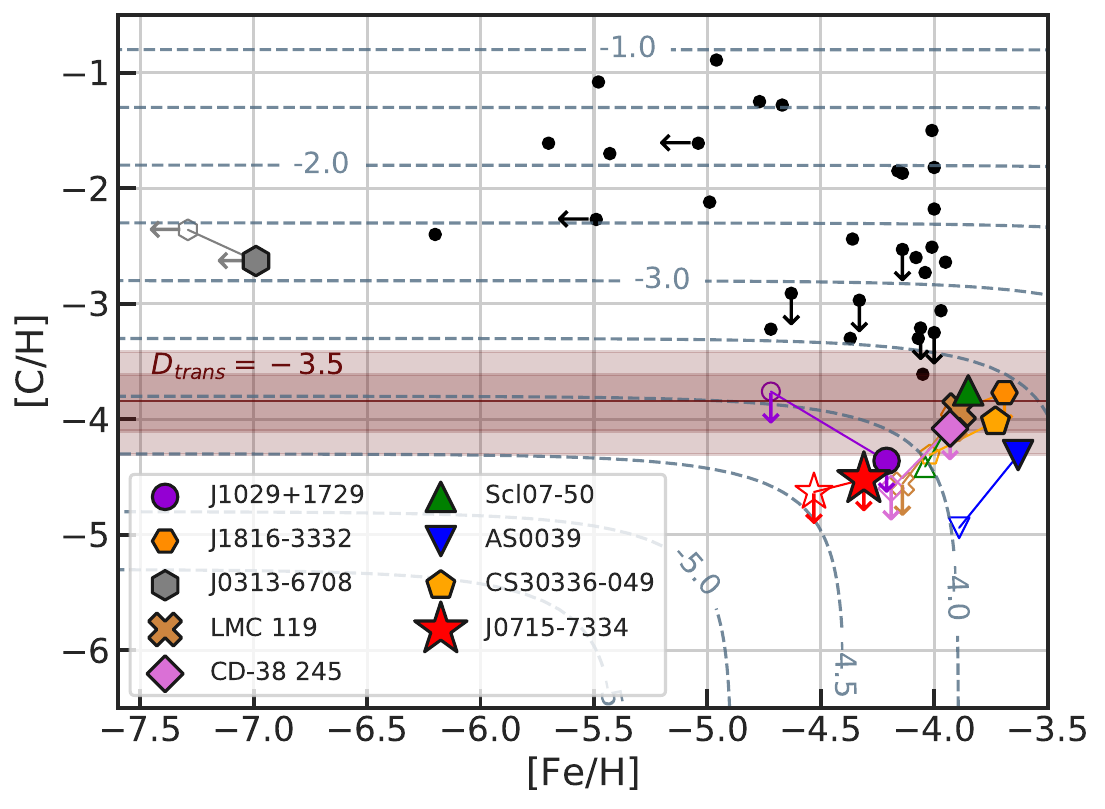}
\vspace{-4mm}
\caption{\textbf{Carbon and iron abundances of ultra-metal-poor stars.}
\umpstar is shown as a large red star. Black points show a literature sample\cite{Abohalima2018,Sestito2019}. Colored points highlight eight other notable stars, with 1D LTE abundances shown as small open symbols and a combination of 1D NLTE, 3D LTE, and 3D NLTE analyses shown as large solid colored symbols (see Methods). The dashed blue contours indicate an approximate total metallicity assuming $\mbox{[Mg/Fe]}=+0.4$ (see Methods). A horizontal dark red line indicates the critical $D_{\rm trans}$ threshold\cite{Bromm2003,Frebel2007} for atomic fine structure line cooling assuming $\mbox{[C/O]} = -0.6$\cite{Amarsi2019}, and the two shaded regions show $-1 < \mbox{[C/O]} < 0$ and an extra 0.2 dex theoretical uncertainty\cite{Bromm2003}.
\label{fig:cfe}
}
\end{figure}

Figure~\ref{fig:cfe} shows \umpstar's iron and carbon abundance compared to 38 literature stars with $\feh_{\rm LTE} \lesssim -4$ that have detailed chemical abundance measurements. 
Approximate total metallicity contours are shown as dashed blue lines (see Methods). 
We highlight eight of the stars with the lowest total metallicities as colored points\cite{Cayrel2004,Lai2008,Caffau2011,Keller2014,Howes2015,Simon2015,Skuladottir2024,Chiti2024}.
We find that \umpstar has the lowest metallicity upper limit known, with $Z < 7.8 \times 10^{-7}$, or $\logzzsun < -4.3$ using $Z_\odot = 0.016$\cite{Lodders2025} after including NLTE, 3D, and evolutionary effects.
This is about two times more metal-poor than the metallicity upper limit claimed in the literature for the previous record holder, {\caffaustar}\cite{Caffau2024} ($Z < 1.66 \times 10^{-6}$). 
Our recalculation of this star's metallicity using 3D NLTE results\cite{Lagae2023} along with the highest S/N 3D LTE carbon upper limit\cite{Caffau2024} gives it a lower metallicity limit ($Z < 9.5 \times 10^{-7}$, see Methods).
\umpstar is over ten times more metal-poor than the most iron-poor star known \kellerstar ($\mbox{[Fe/H]}~{<}-7.0$, $\mbox{[C/H]}=-2.55$){\cite{Nordlander2017}}, emphasizing that iron does not track total metallicity at low [Fe/H].
The exact metallicity of \umpstar depends on assumptions for unmeasured elements, which dominates the uncertainty. 
The most important missing elements are nitrogen and oxygen, where we have unconstraining limits from the NH band (\ref{fig:nhsynth}) and forbidden 6300{\AA} oxygen line.
The numbers above assume that the missing elements are solar-scaled with $\mbox{[X/Fe]}=0$, but other assumptions result in metallicities ranging from $(6.7-21.9) \times 10^{-7}$ (or $-4.38 < \logzzsun < -3.86$, see Methods). 
In almost all cases, \umpstar still has the lowest metallicity upper limit if other stars' metallicities are recomputed using the same assumptions.
Similarly, \umpstar remains the star with the most metal-poor upper limit known if 1D LTE abundances are assumed for all stars ($Z_{\rm 1D,LTE} < 5.0 \times 10^{-7}$).

\begin{figure}[h!]
    \centering
    \includegraphics[width=0.95\linewidth]{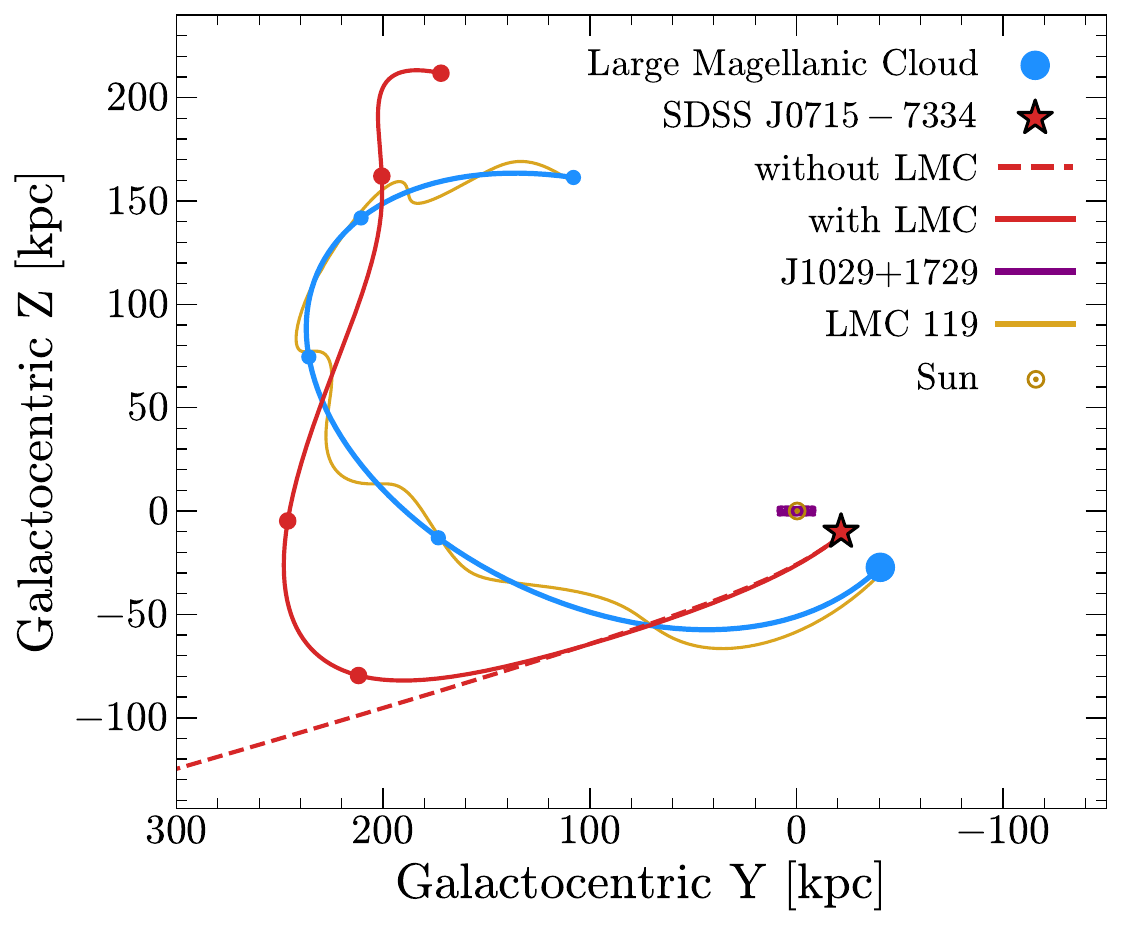}
    \includegraphics[width=0.95\linewidth]{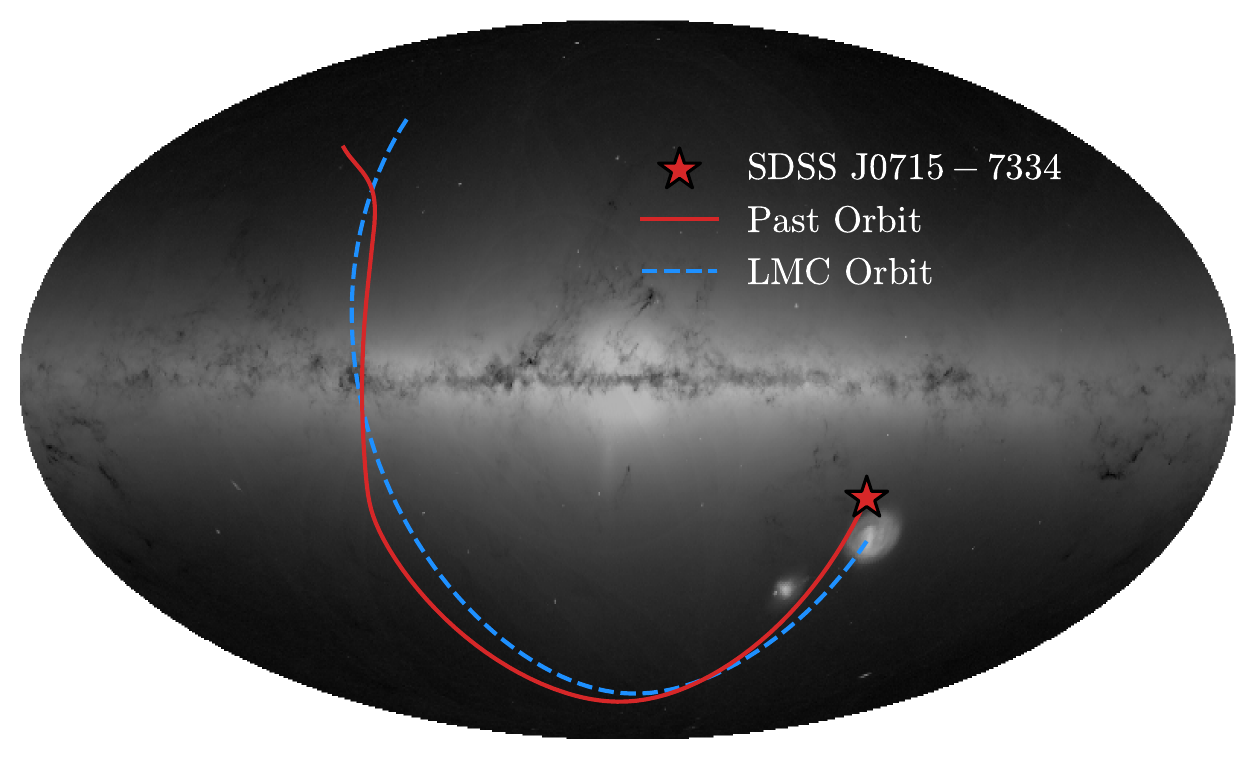}
    \caption{\textbf{Kinematic properties.}
    \textit{Top:}
    Past orbit of \umpstar over $4$~Gyr in Galactocentric coordinates, integrated in a potential that includes the gravitational influence of the LMC (solid lines). 
    Circular markers are placed every $1$~Gyr. 
    For comparison, the past orbit of the LMC itself is shown, along with orbits of the stars \caffaustar (confined to the disk) and LMC-119 (closely bound to the LMC).
    The dashed red line shows the unbound orbit of \umpstar in a MW-only potential.
    \textit{Bottom:} The past orbit of \umpstar and the LMC in Galactic coordinates on-sky, overlaid on the distribution of all stars observed by \textit{Gaia}\cite{gai21}. 
    }
    \label{fig:kinematics}
    \vspace{-2mm}
\end{figure}

We measure a radial velocity from MIKE of $+427.2 \pm 0.9$ km/s and infer a spectrophotometric distance of $26.1^{+1.5}_{-1.6}$\,kpc (see Methods). 
\umpstar initially appeared to be unbound from the Milky Way (dashed red line in the top panel of Figure~3), but its angular momentum and energy is similar to that of the Magellanic Clouds, suggesting a possible association (see Methods and \ref{fig:lxlz}). 
We thus integrated the orbit of \umpstar in a combined time-evolving potential including the Milky Way and Large Magellanic Cloud (LMC, see Methods). 
A typical orbit is shown in Figure~\ref{fig:kinematics}, which follows the LMC as it falls into the Milky Way. 
This realization has a long-period high-eccentricity orbit around the LMC, but more circular and shorter period orbits around the LMC are also allowed given the observational uncertainties.
Some past orbital realizations are not bound to the LMC, but in this case they are unbound to the Milky Way as well, so we do not consider them to be plausible solutions (see Methods and \ref{fig:lmcdistsamples}). 
\umpstar is thus a galactic immigrant, born in or near the LMC and recently captured by the Milky Way. 
This contrasts with \caffaustar, which has a Milky Way thick disk orbit\cite{Sestito2019,Caffau2024} and was thus likely born in the Milky Way itself. 
The other known star in the LMC with $\mbox{[Fe/H]}_{\rm LTE}<-4$, LMC-119\cite{Chiti2024}, is much more tightly bound to the LMC (Fig 3, solid yellow line). 
LMC-119 is also not carbon enhanced, in contrast to the dominant population in the Milky Way halo\cite{Christlieb2002,Frebel2005,Frebel2015} and the lowest mass ultra-faint dwarf galaxies\cite{Ji2020,Chiti2025}, suggesting that carbon enhancement may be environment dependent\cite{Chiti2024}.

It is a long-standing question what physics causes the transition from metal-free star formation with a top-heavy initial mass function to the present-day initial mass function\cite{Bromm2009,Frebel2015,Klessen2023}. 
One idea is that metals change the thermal evolution of gas, causing fragmentation primarily either through carbon and oxygen fine structure line cooling or dust thermal cooling\cite{Schneider2003,Schneider2012,Bromm2003}.
The minimum ``critical metallicity'' to allow low mass star formation is $\logzzsun \gtrsim -3.5$ for atomic fine structure cooling\cite{Bromm2003,Frebel2007} or $\logzzsun \gtrsim -5$ for dust thermal cooling\cite{Schneider2003,Schneider2012,Schneider2012b,Chiaki2017}. 
The critical threshold to activate fine structure cooling is often written as $D_{\rm trans}=\log_{10}(10^{\mbox{[C/H]}} + 0.3 \times 10^{\mbox{[O/H]}}) > -3.5$\cite{Frebel2007}.
Previously, \caffaustar was the only star known with $D_{\rm trans} < -3.5$\cite{Caffau2011,Caffau2024}.
Limberg et al.\cite{Limberg2025} suggested that \umpstar could also be below the $D_{\rm trans}$ threshold, but their carbon upper limit was too high to be certain. 
Our stringent carbon upper limit now gives $D_{\rm trans} < -4.2_{-0.2}^{+0.3}$, where the dominant uncertainty is the assumed C/O ratio $-1 < \mbox{[C/O]} < 0$. 
Thus, \umpstar is now the second known star that can not form through atomic fine structure cooling (Fig~\ref{fig:cfe}), proving that dust cooling is required and also operates in environments beyond the Milky Way.
We estimate that ${>}1\%$ of metals must be depleted into dust in order to explain the formation of \umpstar, possibly requiring dust grain growth during its protostellar phase (see Methods).

\begin{figure}[h!]
    \centering
    \includegraphics[width=\linewidth]{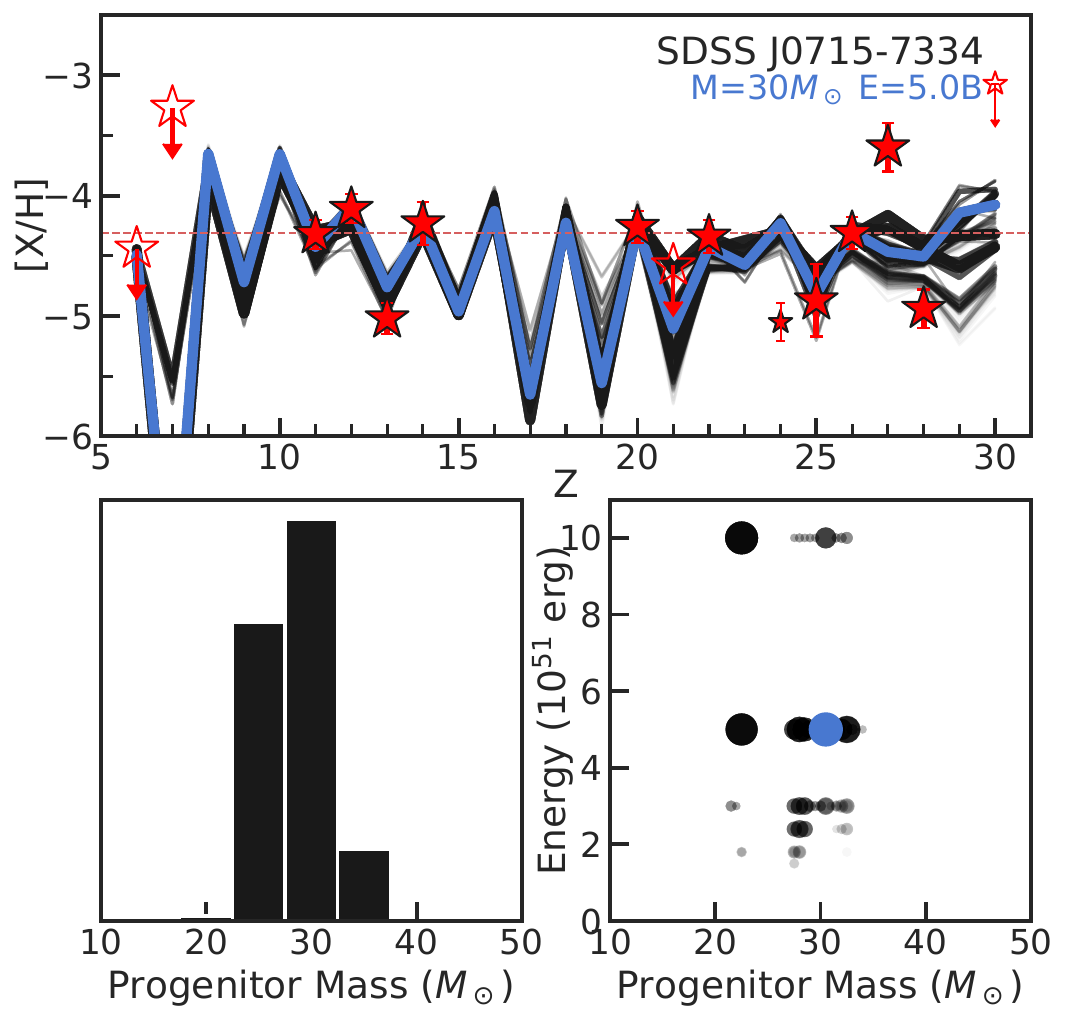}
    \caption{\textbf{Population~III Supernova Progenitor Constraints.}
    \textit{Top:} Chemical abundance pattern of \umpstar. Filled red stars show measured abundances with $1\sigma$ uncertainties; open red stars with arrows are upper limits (treated as hard cutoffs). Sc is treated as an upper limit due to model uncertainties; Cr and Zn are excluded\cite{Heger2010}. The blue line and caption shows the best-fit model; black lines show other models within $95\%$ confidence, with opacity indicating fit quality.
    \textit{Bottom left:} Progenitor mass distribution of models weighted using total absolute error (see Methods).
    \textit{Bottom right:} Explosion energy vs. progenitor mass for the same models; point size scales with fit quality. Best-fit model shown in blue.
    }
    \label{fig:popiii}
\end{figure}

The detailed chemical abundances of the most metal-poor stars can be linked back to the properties of metal-free Population~III stars through supernova nucleosynthesis models\cite{Bromm2009,Frebel2015,Klessen2023}. 
\umpstar is an especially clean probe of Population~III, as its distant halo orbit completely precludes significant surface contamination from the interstellar medium\cite{Frebel2009,Shen2017} and its large convective envelope removes any diffusive settling effects. 
Figure~\ref{fig:popiii} shows the result of a fit to the detailed elemental composition of \umpstar using theoretical metal-free supernova yields\cite{Heger2010} (see Methods).
Stringent upper limits are placed on the neutron-capture elements (\ref{fig:boxplot}) but not used in the fit.
Overall, this star is best explained by a $30~M_\odot$ progenitor star and a high explosion energy around $5 \times 10^{51}$ erg.
A weighted average and standard deviation gives $M_{\rm prog} = 27.0 \pm 3.9 M_\odot$ and $E_{\rm expl} = 6.0 \pm 2.6 \times 10^{51}$\,erg (see Methods).
Leave-one-element-out tests show that the initial mass fit is driven by the C upper limit (ruling out models around $10-15 M_\odot$) and Na and Mg (shifting the best fit mass up from 25$M_\odot$ to 30$M_\odot$). 
Elements affecting the explosion energy are Ca and Mn (ruling out the $10^{52}$ erg models) and Co (preferring models $>3 \times 10^{51}$ erg).

As the two most metal-poor stars known, it is worth comparing the Population~III progenitors for \umpstar with \caffaustar. 
\caffaustar has a Population~III progenitor with low initial mass $10-20 M_\odot$\cite{Lagae2023} and a low explosion energy ${\lesssim} 1.5 \times 10^{51}$ erg, while 
\umpstar has a Population~III progenitor with much higher initial mass $25-35 M_\odot$ and a high explosion energy ${\gtrsim} 5 \times 10^{51}$ erg.
These two stars originate in clearly different environments.
We speculate that this could be an early indication of environment dependence in Population~III star formation, a scenario known as Population~III.2\cite{Johnson2006,McKee2008}.
The LMC forms several comoving megaparsecs away from the proto-Milky Way\cite{Chiti2024}, close enough for radiative feedback but not chemical feedback\cite{Madau2001}. 
Thus, its Population~III star formation could be affected by Lyman Werner or ionizing radiation from the proto-Milky Way, which in turn may change the Population~III initial mass function\cite{Bromm2009,Clark2011,Klessen2023}.
However, many more similarly metal-poor stars will need to be found in different environments to test this hypothesis.

The launch of the James Webb Space Telescope has led to a flurry of discoveries of extremely metal-poor high-redshift galaxies\cite{Fujimoto2025,Nakajima2025,Morishita2025}.
The metallicity upper limits have been claimed to be as low as $\mbox{[Z/H]} < -3$.
These are undoubtedly exciting objects, but the metallicity constraints remain an order of magnitude away from legitimate claims of detecting Population~III stars.
The lowest [OIII]/H$\beta$ ratio measured so far is $<0.22$\cite{Morishita2025}, but a galaxy with the same metallicity as \umpstar would have [OIII]/H$\beta$ $< 0.01$\cite{Katz2023}.
Thus, at least 10 times better signal-to-noise ratios are needed in order to show that these high-redshift galaxies are not Population~II galaxies composed of stars like \umpstar.
The search for Population~III stars continues.

\begin{table*}
\centering
\setlength{\tabcolsep}{2.5pt}
\caption{Chemical Abundances of \umpstar\label{tab:abunds}}
\begin{tabular}{lrrrrrrrcc}
\hline
Elem & N & A(X)   & A(X)   & A(X)   &  [X/H] & [X/Fe] & $\Delta$A(X) & $\sigma$ & $\sigma$ \\
     &   & Solar  & LTE    & NLTE   &  NLTE  & NLTE   &              & [X/H]    & [X/Fe]   \\
\hline
Li I  &  1 &$ 3.30$&${<}{ 0.33}$&${<}{ 0.33}$&${<}{-2.97}$&${<}{ 1.33}$&       &$    $&$    $\\
C-H   &  1 &$ 8.51$&${<}{ 3.88}$&${<}{ 3.99}$&${<}{-4.52}$&${<}{-0.22}$&$+0.11$&$    $&$    $\\
N-H   &  1 &$ 7.94$&${<}{ 4.56}$&${<}{ 4.56}$&${<}{-3.38}$&${<}{ 0.92}$&       &$    $&$    $\\
O I   &  1 &$ 8.76$&${<}{ 6.91}$&${<}{ 6.91}$&${<}{-1.85}$&${<}{ 2.45}$&       &$    $&$    $\\
Na I  &  2 &$ 6.30$&$   { 1.86}$&$   { 1.92}$&$   {-4.38}$&$   {-0.07}$&$+0.06$&$0.12$&$0.09$\\
Mg I  &  6 &$ 7.57$&$   { 3.23}$&$   { 3.49}$&$   {-4.08}$&$   { 0.22}$&$+0.26$&$0.12$&$0.11$\\
Al I  &  2 &$ 6.46$&$   { 1.12}$&$   { 1.43}$&$   {-5.03}$&$   {-0.73}$&$+0.31$&$0.13$&$0.09$\\
Si I  &  1 &$ 7.55$&$   { 3.04}$&$   { 3.28}$&$   {-4.27}$&$   { 0.03}$&$+0.24$&$0.18$&$0.13$\\
K I   &  1 &$ 5.12$&${<}{ 2.18}$&${<}{ 2.18}$&${<}{-2.94}$&${<}{ 1.37}$&       &$    $&$    $\\
Ca I  &  2 &$ 6.33$&$   { 1.87}$&$   { 2.08}$&$   {-4.25}$&$   { 0.06}$&$+0.21$&$0.13$&$0.10$\\
Ca II &  2 &$ 6.33$&$   { 2.05}$&$   { 2.07}$&$   {-4.26}$&$   { 0.04}$&$+0.02$&$0.24$&$0.17$\\
Sc II &  3 &$ 3.08$&$   {-1.43}$&$   {-1.43}$&$   {-4.51}$&$   {-0.21}$&       &$0.21$&$0.13$\\
Ti II & 13 &$ 4.94$&$   { 0.29}$&$   { 0.62}$&$   {-4.33}$&$   {-0.02}$&$+0.33$&$0.14$&$0.07$\\
Cr I  &  2 &$ 5.67$&$   { 0.57}$&$   { 0.60}$&$   {-5.08}$&$   {-0.77}$&$+0.03$&$0.16$&$0.11$\\
Mn I  &  1 &$ 5.52$&$   {-0.12}$&$   { 0.56}$&$   {-4.96}$&$   {-0.65}$&$+0.68$&$0.30$&$0.27$\\
Fe I  & 58 &$ 7.49$&$   { 2.97}$&$   { 3.19}$&$   {-4.30}$&$   { 0.00}$&$+0.22$&$0.13$&$    $\\
Fe II &  6 &$ 7.49$&$   { 2.76}$&$   { 2.76}$&$   {-4.73}$&$   {-0.42}$&$-0.00$&$0.15$&$    $\\
Co I  &  2 &$ 4.91$&$   { 0.51}$&$   { 1.39}$&$   {-3.52}$&$   { 0.79}$&$+0.88$&$0.20$&$0.14$\\
Ni I  &  6 &$ 6.25$&$   { 1.28}$&$   { 1.28}$&$   {-4.97}$&$   {-0.66}$&$-0.00$&$0.16$&$0.08$\\
Zn I  &  1 &$ 4.65$&${<}{ 1.49}$&${<}{ 1.49}$&${<}{-3.16}$&${<}{ 1.15}$&       &$    $&$    $\\
Sr II &  1 &$ 2.93$&${<}{-3.35}$&${<}{-3.35}$&${<}{-6.28}$&${<}{-1.97}$&       &$    $&$    $\\
Ba II &  1 &$ 2.22$&${<}{-3.31}$&${<}{-3.31}$&${<}{-5.53}$&${<}{-1.22}$&       &$    $&$    $\\
\hline
\end{tabular}

N is the number of lines used.
The A(X) NLTE, [X/H], and [X/Fe] columns are provided in 1D NLTE when possible, using the correction in the $\Delta$A(X) column.
The correction for C-H is not using NLTE, but instead calculated as the difference between the 1D LTE analysis and the combination of a 3D correction of $-0.45$ dex\cite{Eitner2024} and an evolutionary correction of $+0.56$ dex\cite{Placco2014}.
Solar abundances adopted are from Lodders et al.\cite{Lodders2025,Bergemann2025}.
[X/Fe] is computed using [Fe\,I/H].
Abundance uncertainties on [X/H] and [X/Fe] differ because they consistently propagate correlations with respect to stellar parameters.
\end{table*}

\newpage

\begin{center}
{\bf \Large \uppercase{Methods} }
\end{center}

\captionsetup[figure]{labelformat=empty}
\captionsetup[table]{labelformat=empty}
\setcounter{figure}{0}    
\renewcommand{\thefigure}{Extended Data Figure \arabic{figure}}
\renewcommand{\thetable}{Extended Data Table \arabic{table}}

\noindent
{\bf Observations}
\\
\noindent
The fifth-generation Sloan Digital Sky Survey (SDSS-V\cite{Kollmeier2026}) has been executing an all-sky survey of the MW halo since 2021, using twin Baryon Oscillation Spectroscopic Survey spectrographs (BOSS\cite{Smee2013}) at the Apache Point Observatory (APO) and Las Campanas Observatory (LCO). 
These halo stars are targeted using a variety of spectro-photometric selections that prioritize distant and metal-poor stars\cite{Chandra2025}. 
\umpstarsdss{} was observed by BOSS at LCO on 2024 Dec 16 for a single 15-minute exposure, which achieved signal-to-noise ratios ${\approx}~25$ per pixel.
The coordinates of the star are $(\alpha,\delta) = (108.91087^\circ, -73.58141^\circ)$, the \textit{Gaia} source ID is \umpstargaia, the 2MASS ID is \umpstartmass, and the SDSS ID is \umpstarsdssid.
% and the first time this star was cataloged was as USNO-A1.0 0150-04773636. 
The star was targeted for belonging to the infrared color-selected metal-poor candidate selection of Schlaufman \& Casey\cite{Schlaufman2014} and a distant red giant selection\cite{Conroy2018, Conroy2021, Chandra2023b, Chandra2025}. 
Stellar parameters were determined using the \code{MINESweeper} pipeline of the SDSS-V halo survey\cite{Cargile2020, Chandra2025}. 
The pipeline estimated a metallicity [Fe/H]~$\approx -3.8$ for \umpstarsdss{}, near the lowest edge of the allowable parameter grid and the lowest metallicity of stars observed in SDSS-V to date. 
Visual inspection showed the star was clearly near the tip of the RGB, but no metal absorption lines beyond a weak Calcium K line were detected in the BOSS spectrum, flagging \umpstarsdss{} for high-resolution spectroscopic observations. 

We obtained 225 minutes of high-resolution followup spectroscopy of \umpstar on 2025 Mar 21-22 with the MIKE spectrograph\cite{Bernstein2003} on the Magellan/Clay telescope.
We used the 0.7 arcsecond slit, which provides resolution 35,000 in the blue and 28,000 in the red (split at ${\approx}~5000${\AA}).
The data were reduced with CarPy\cite{Kelson2003}, obtaining signal-to-noise ratios (50, 85, 165)/pixel at (4000{\AA}, 4500{\AA}, 6500{\AA}) respectively, with about two pixels per resolution element.
The heliocentric radial velocity was $427.2 \pm 0.9 \kms$, where the uncertainty is calculated using the order-to-order scatter that quantifies the dominant wavelength calibration uncertainty\cite{Ji2020}.
The velocity differs from the SDSS-V measurement by ${\approx}2\sigma$, but systematic wavelength calibration uncertainties of ${\approx}$~10 km s$^{-1}$ are known in SDSS-V\cite{Chandra2025} so this is not evidence for binarity.
Spectra were normalized and stitched using the \code{LESSPayne} environment\cite{Ji2025ascl},
which combines \textit{The Payne}\cite{Ting2019} with the \code{smhr} graphical user interface\cite{Casey2014}.

\vspace{1mm}
\noindent
{\bf Stellar Parameter Analysis}
\\
\noindent
We adopt an effective temperature $\Teff = 4700 \pm 100$\,K and surface gravity $\logg = 1.10 \pm 0.25$ based on a consideration of photometry, spectroscopy, and asteroseismology as detailed in the following section.
The resulting metallicity is $\feh = -4.53 \pm 0.2$, assuming 1D Local Thermodynamic Equilibrium (LTE).
This star is at the extreme end of essentially all model grids considered, but this range covers what we consider to be all reasonable solutions.

Photometric stellar parameters were determined in five ways and obtained $\Teff = 4650-4800$\,K and $\logg \approx 1.1$.
First, we used six color-temperature relations\cite{Mucciarelli2021} based on Gaia DR3 $G$, $BP$, $RP$ photometry; 2MASS $K_s$ photometry; and the Schlafly \& Finkbeiner\cite{Schlafly2011} extinction of $A_V = 0.50$ to obtain $\Teff = 4668 \pm 16$\,K.
Next, we derived color-temperature and color-gravity relations from the most metal-poor isochrones available for the MIST, Dartmouth, PARSEC, and BaSTI isochrones considering the same six colors and extinction, resulting in $\Teff = 4678 \pm 36$\,K and $\logg = 1.06 \pm 0.15$. Using the larger Schlegel et al.\cite{Schlegel1998} reddening $A_V = 0.58$ resulted in $\Teff = 4726 \pm 38$\,K and $\logg=1.17\pm0.15$.
Third, we used the \code{isochrones} package\cite{mor15} with \code{MultiNest}{\cite{fer08}} and MIST isochrones\cite{Choi2016} to fit the
zero point-corrected Gaia DR3 parallax\cite{gai21,fab21,lin21a,lin21b,row21,tor21} and photometry from
GALEX NUV\cite{bia17};
SkyMapper DR4 uvgriz\cite{onk24}; 
Gaia DR3 $G$\cite{gai16,gai21,fab21,rie21,row21,tor21};
Two-micron All-sky Survey (2MASS) $JHK_{\text{s}}$\cite{skr06};
and Wide-field Infrared Survey Explorer CatWISE2020 $W1W2W3$\cite{wri10,mai11,eis20,mar21}.
We adopt a uniform extinction prior $0 < A(V) < 1$ mag; a distance prior based on a parallax-only geometric distance\cite{bai21} but increasing the uncertainty by a factor of 5; and a log-uniform age prior from 1 Gyr to 13.8 Gyr.
The MIST isochrones do not allow \feh below $-4$, so we put a uniform prior $-4.0 < \feh < -3.5$.
This results in stellar parameters $\Teff = 4780_{-20}^{+20}$\,K, $\log g = 1.04_{-0.05}^{+0.06}$ with reddening $A_V = 0.65^{+0.04}_{-0.03}$, corresponding to a distance $d = 25.6_{-1.7}^{+1.3}$\,kpc.
Inspection of the results suggested they could potentially be affected by finite sampling of the isochrone near the tip of the red giant branch (TRGB), so we ran another test using a more finely sampled MIST isochrone fixing $\feh=-4$ and using only redder photometry like Gaia RP, 2MASS $JHK_{\rm s}$, and WISE $W1W2$, which resulted in essentially the same result of $\Teff=4787$\,K, $\logg=1.1$, $d=24.3$\,kpc with $A_V=0.58$.
Finally, for an independent atmosphere model, we used the PHOENIX stellar atmospheres\cite{Husser2013} to fit the broadband SED using 2MASS $JHK_{\rm s}$; WISE $W1W2W3$; Gaia G, BP, RP, and XP spectrophotometry; and GALEX NUV, resulting in $\Teff = 4750 \pm 50$\,K with $A_V = 0.625$\cite{Stassun2017,Stassun2018}.

Spectroscopic stellar parameters were determined in three ways and obtained $\Teff = 4500-4600$\,K.
First, we performed a 1D LTE analysis with an empirical temperature correction\cite{Frebel2013}.
We measured equivalent widths of Fe\,I and Fe\,II lines, then used \code{MOOG} with scattering\cite{Sneden1973,Sobeck2011,Sneden2012} and ATLAS model atmospheres\cite{Kurucz1979,Castelli2003} to ensure no trend in Fe line abundances with respect to excitation potential, ionization state, and reduced equivalent width.
We then applied the empirical temperature correction and rebalanced the Fe lines holding the temperature fixed, resulting in $\Teff=4540$\,K, $\logg=0.95$, $\feh=-4.77$, and microturbulence $\xi_t = 1.62\kms$.
We also used \code{TSFitPy}{\cite{Gerber2023}}, which uses \code{Turbospectrum}{\cite{Plez2012}} and the standard \code{MARCS} model atmospheres\cite{Gustafsson2008} along with precomputed non-LTE (NLTE) departure coefficients for Fe\cite{Bergemann2012b} from \code{MULTI}{\cite{Carlsson1986}}, and attempted to find a NLTE spectroscopic balance. Our solution reached the edge of the departure coefficient grid at $\Teff = 4500$\,K, but the final solution would need to have a lower temperature. Fixing $\Teff=4500$\,K, we found $\logg=1.4$, $\feh=-4.65$, and $\xi_t=1.6\kms$ in 1D NLTE. Using a lower temperature would further lower \logg and the metallicity.
Finally, we ran a fit to the H$\beta$ Balmer line profile. We used \code{Korg}\cite{Wheeler2023,Wheeler2024} to generate the hydrogen line profiles, then simultaneously fit the temperature, surface gravity, metallicity, and continuum\cite{Casey2017}.
The results suggested a temperature $\Teff \approx 4600$\,K, with uncertainties driven primarily by how much of the H core was masked.
Note that we could not use the H$\alpha$ Balmer line profile in this star as it shows substantial emission features associated with mass loss at the tip of the red giant branch.
There are also small emission features in the Ca HK line cores that may also suggest some level of chromospheric activity in this star.

Finally, we put a constraint on \logg with asteroseismology using photometry from the \textit{TESS} mission\cite{Ricker2015_TESS}.
\umpstar is located in the southern continuous viewing zone of \textit{TESS}, which currently has a total of $2.8$ years of observations.
We detrended the time-series photometry by\cite{Stello2022}: 
downloading TESS photometry from the Mikulski Archive for Space Telescopes (MAST) database using the \texttt{tesscut} python package\cite{Brasseur2019};
% \footnote{\textit{TESS} data can be found here: \href{https://archive.stsci.edu/doi/resolve/resolve.html?doi=10.17909/t9-nmc8-f686}{ doi:10.17909/t9-nmc8-f686}} , 
manually defining aperture masks in the \textit{TESS} full frame images;
using the regression corrector function in \texttt{LightKurve}\cite{LightKurve} to remove systematic background trends for each sector separately;
stitching light curves together without filling large gaps; and 
computing the Lomb-Scargle periodogram\cite{Lomb1976,Scargle1982} of the full light curve to get the power spectrum. 
To reduce the noise in the final stitched light curve, we omitted sectors with light curves that displayed long-term trends in the photometry after the regression corrector step. In total, we used 27 sectors of \textit{TESS} data, equating to an observational duty cycle of 2 years. Despite the considerable length of the observations, the resulting asteroseismic signal had a low signal-to-noise ratio, and we treat it as a marginal detection. This can be attributed to the star’s faint magnitude, which is at the detectability limit of the \textit{TESS} instrument.
Using the asteroseismic pipeline \texttt{pyMON}\cite{Howell2025}, a tentative measurement for the frequency of the maximum acoustic power was found to be $\nu_{\rm max} = 18\mu$Hz. 
This corresponds to a surface gravity of $\logg_{\rm seismic} = 2.2$ using the seismic scaling relation\cite{KjeldsenBedding1995}: $\ensuremath{g}_{\rm seismic}\propto\nu_{\rm max} \ensuremath{T_{\mathrm{eff}}}^{0.5}$ . 
There are known deviations from the seismic scaling relations for metal-poor stars, where masses larger than expected are estimated\cite{Epstein14_metal_poor_seismo, Schonhut-Stasik2024}. 
To reconcile this, a scaling factor to the $\nu_{\rm max}$ quantity is implemented in the seismic scaling relations, consequently reducing the seismic mass and $\logg_{\rm seismic}$ estimates\cite{Huber2024,Larsen2025}. Calibrations of the $\nu_{\rm max}$ scaling factor require detailed modeling of the individual frequency peaks, which is unfeasible for \umpstar given the low signal-to-noise ratio of the asteroseismic signal. Therefore, we conclude that lower $\logg$ values cannot be ruled out with asteroseismology and thus adopt $\logg_{\rm seismic} = 2.2$ as an upper bound.
This still excludes the higher $\logg$ spectroscopic solutions, which could be biased due to the small number of Fe\,II lines available.

It is clear that systematic modeling choices for such an extreme cool and metal-poor star dominate the overall stellar parameter uncertainty.
We thus decide to adopt an intermediate value $\Teff=4700 \pm 100$\,K. Putting this temperature range into the MIST, Dartmouth, PARSEC, and BaSTI isochrones results in $\logg = 1.1 \pm 0.25$.
These $\Teff$ and $\logg$ ranges cover essentially all values in the exploration above.
Using these stellar parameters, balancing Fe\,I abundances with respect to reduced equivalent width in 1D LTE with MOOG and ATLAS results in $\xi_t = 1.60\kms$, and we adopt a 0.3\kms systematic uncertainty as many $\logg-\xi_t$ relations predict $\xi_t \approx 2.0$ at such low \logg values\cite{Ji2023}.
We adopt a model metallicity $\mbox{[M/H]} = -4.53 \pm 0.15$, and for our fiducial ATLAS analysis we use $\alphafe = 0.0$, though adopting $\alphafe = +0.4$ results in nearly identical abundances.
The stellar parameter uncertainties will dominate the chemical abundance uncertainties.

The final choice of stellar parameters results in a significant ionization imbalance between $\mbox{[Fe\,I/H]}=-4.30$ and $\mbox{[Fe\,II/H]}=-4.73$, a 2.1$\sigma$ difference given the quoted uncertainties.
This is a common occurrence in the analysis of the most metal-poor red giant stars with $\feh \lesssim -3$\cite{Karovicova2020,Martin2022,Caffau2025}.
One suggested possibility is that the Fe\,I NLTE corrections are too large at such low \feh\cite{Caffau2024,Caffau2025}, even though they are clearly more accurate at higher metallicities $\feh \gtrsim -2.5$\cite{Bergemann2017}.
We also verified that similar or larger NLTE corrections are consistently found across many independent calculations with different atomic data and model atmospheres \cite{Bergemann2012b,Lind2012,Mashonkina2017,Ezzeddine2018}.
Another possibility is that our adopted surface gravity from isochrones is too low. This would require the stellar evolution models to be incorrect, which for example could occur if the mixing length varies substantially between $\feh = -2.5$ where the isochrones have been empirically checked against globular clusters and $\feh < -4$\cite{Tayar2017,Choi2018}. We investigated several metal-poor red giants with Gaia DR3 parallaxes and compared our purely photometric \logg derivation with that using the fundamental Stefan-Boltzmann law, finding offsets within 0.1 dex. In particular, \cdstar is a $\feh=-4$ red giant with a parallax signal-to-noise ratio ${>}14$ in Gaia DR3. We photometrically derive $\logg = 1.48 \pm 0.19$, while using geometric distances\cite{bai21} with color-temperature relations\cite{Mucciarelli2021} and bolometric corrections\cite{Casagrande2018} obtains $\logg = 1.58 \pm 0.07$.
Thus there is no clear solution to the ionization imbalance, reflecting the current state of the literature.

One option to avoid the question of ionization imbalance would be to adopt the Fe\,II abundance as the metallicity for this star, as is often done for metal-poor red giants\cite{Ji2020b,Martin2022}. We decided to keep the NLTE-corrected Fe\,I abundance as our fiducial value as most of the stars in this metallicity range do not have Fe\,II lines, and this also produces a more conservative upper limit on the metallicity of \umpstar ($\logzzsun<-4.3$ instead of $\logzzsun<-4.7$).

\vspace{1mm}
\noindent
{\bf Chemical Abundance Analysis}
\\
\noindent
We start with a standard 1D LTE chemical abundance analysis on all elements using \code{LESSPayne}{\cite{Ji2025ascl}} and \code{TSFitPy}{\cite{Gerber2023}}.
The 1D LTE \code{LESSPayne} analysis uses \code{MOOG} with scattering and interpolates ATLAS model atmospheres from a precomputed grid.
A new linelist was constructed by identifying all visible absorption lines in the spectrum, then adopting atomic data from {\code{linemake}}\cite{Placco2021}.
This uses laboratory data when possible, and Kurucz atomic data when not possible.
Nearly all element abundances were determined using equivalent widths fit using Gaussian profiles in \code{LESSPayne}.
This includes species like Al\,I and Si\,I that would normally require syntheses in iron-poor stars due to large carbon enhancements, and we verified there were no blends in these lines.
However, we synthesized three iron-peak species (Sc\,II, Mn\,I, Co\,I) due to hyperfine structure splitting.
Additionally, for non-detections of Li\,I, O\,I, K\,I, Zn\,I, Sr\,II, and Ba\,II, we determined upper limits by finding the best-fit model of the continuum and abundance simultaneously, then increasing the abundance until $\Delta\chi^2=25$, close to a formal $5\sigma$ upper limit\cite{Ji2020b}. C and N constraints are based on molecular bands that require a more careful continuum treatment and discussed later.

To check systematic uncertainties on the model atmosphere, line list, and radiative transfer code, we also conducted a 1D LTE abundance analysis using \code{TSFitPy}{\cite{Gerber2023}}, which wraps \code{Turbospectrum}{\cite{Plez2012}} and uses the standard MARCS model atmospheres\cite{Gustafsson2008}.
The line list was from Gaia-ESO\cite{Heiter2021} and extended with VALD\cite{VALD} down to $\lambda > 3700${\AA}, so the \code{TSFitPy} analysis is restricted to redder lines.
We iteratively synthesized synthetic spectra in LTE and calculated their equivalent widths, matching them to the measured equivalent widths.
Differences between the \code{MOOG}/ATLAS analysis and the \code{Turbospectrum}/MARCS analysis are included as a contribution to the per-line systematic uncertainty (\code{dLTE}).

We then determined detailed line-by-line abundance uncertainties using \code{LESSPayne}.
For each spectral line $i$, we calculate statistical, stellar parameter, and systematic uncertainties.
Statistical uncertainties ($e_{\text{stat}, i}$) are obtained by propagating equivalent width uncertainties on Gaussian fits from \code{LESSPayne}, including continuum placement\cite{Ji2020b}.
Stellar parameter uncertainties are calculated by re-determining the abundances in three alternate stellar parameter scenarios: 
cooler ($\Teff=4600$\,K, $\logg=0.85$, $\xi_t=1.60$, $\feh=-4.64$; \code{dSP1}), 
warmer ($\Teff=4800$\,K, $\logg=1.35$, $\xi_t=1.60$, $\feh=-4.41$; \code{dSP2}), and 
$\xi_t$ uncertainty ($\Teff=4700$\,K, $\logg=1.10$, $\xi_t=1.90$, $\feh=-4.53$; \code{dSP3}).
For each line, we then subtract the abundances obtained with the alternate set of stellar parameters from those obtained with the fiducial set of stellar parameters, 
e.g., \code{dSP1}{$_i$} = $\log \epsilon_{i, \text{cool}} - \log \epsilon_{i, \text{fiducial}}$. 
With these abundance differences, we calculate the total stellar parameter uncertainty by taking the quadrature sum $e_{\text{SP},i}^2 = \max(|\code{dSP1}_i|, |\code{dSP2}_i|)^2 + \code{dSP3}_i^2$.
Systematic uncertainties $s_i$ for each line are calculated by taking the largest of three uncertainties: the difference between MOOG/ATLAS and Turbospectrum/MARCS ($\code{dLTE}_i$), a systematic uncertainty such that adding statistical and systematic uncertainties matches the line-to-line standard deviation\cite{Ji2020b}, and a minimum per-line systematic floor of 0.1 dex.

For each chemical species $X$, we calculate the abundance $\log \epsilon(X)$ as an inverse-variance weighted average, $\log\epsilon(X) = \sum_i w_i \log\epsilon_i/\sum_i w_i$.
We adopt per-line weights $w_i = \left[e_{\text{stat},i}^2 + e_{\text{SP},i}^2 + s_i^2\right]^{-2}$ that include the per-line statistical, stellar parameter, and systematic uncertainties.
The stellar parameter uncertainties for species $X$ are also calculated using the weights, e.g. $\code{dSP1}_X = \sum_i w_i \code{dSP1}_i/\sum_i w_i$.
To calculate the total statistical uncertainty, we take $\sigma_{\text{stat},X}^{-2} = \left[\sum_i 1/(e_{\text{stat},i}^2 + s_i^2) \right]$, excluding the stellar parameter uncertainty as this is included differently for [X/H] and [X/Fe].
The uncertainty on $\log\epsilon(X)$ is then given by $\sigma_{X}^2 = \sigma_{\text{stat},X}^2 + \max(|\code{dSP1}_X|, |\code{dSP2}_X|)^2 + \code{dSP3}_X^2$, which is also the uncertainty on [X/H].
The uncertainty on [X/Fe] adds statistical uncertainties for both $X$ and Fe, then accounts for the correlated errors on $X$ and Fe with respect to stellar parameters, i.e. $\sigma_{\text{[X/Fe]}} = \sigma_{\text{stat},X}^2 + \sigma_{\text{stat},\text{Fe}}^2 + \max(|\code{dSP1}_X - \code{dSP1}_{\rm Fe}|, |\code{dSP2}_X - \code{dSP2}_{\rm Fe}|)^2 + (\code{dSP3}_X - \code{dSP3}_{\rm Fe})^2$.
In general, the uncertainties on [X/Fe] are lower as the stellar parameter uncertainties partially cancel out.

Finally, we calculated non-LTE corrections for most elements using \code{TSFitPy}.
Departure coefficients are precomputed from model atoms for
sodium and aluminum\cite{Ezzeddine2018}, 
magnesium\cite{Bergemann2017}, 
silicon\cite{Bergemann2013,EMagg2022}, 
calcium\cite{Mashonkina2017,Semenova2020},  
titanium\cite{Bergemann2011},
chromium\cite{BergemannCescutti2010},
manganese\cite{Bergemann2019},
iron\cite{Bergemann2012b,Semenova2020},
cobalt\cite{Bergemann2010,Yakovleva2020}, 
and nickel\cite{bergemann2021,Voronov2022}.
Only lines from $3700-9200${\AA} are used.
We calculate line-by-line NLTE corrections, given by the column $\code{dNLTE}_i$,
then do the same weighted sum as above to determine a total NLTE correction for each species.
The total corrections are shown in Table~\ref{tab:abunds}.
We adopt the 1D NLTE abundances as our final abundances.

\begin{figure*}
    \centering
    \includegraphics[width=\linewidth]{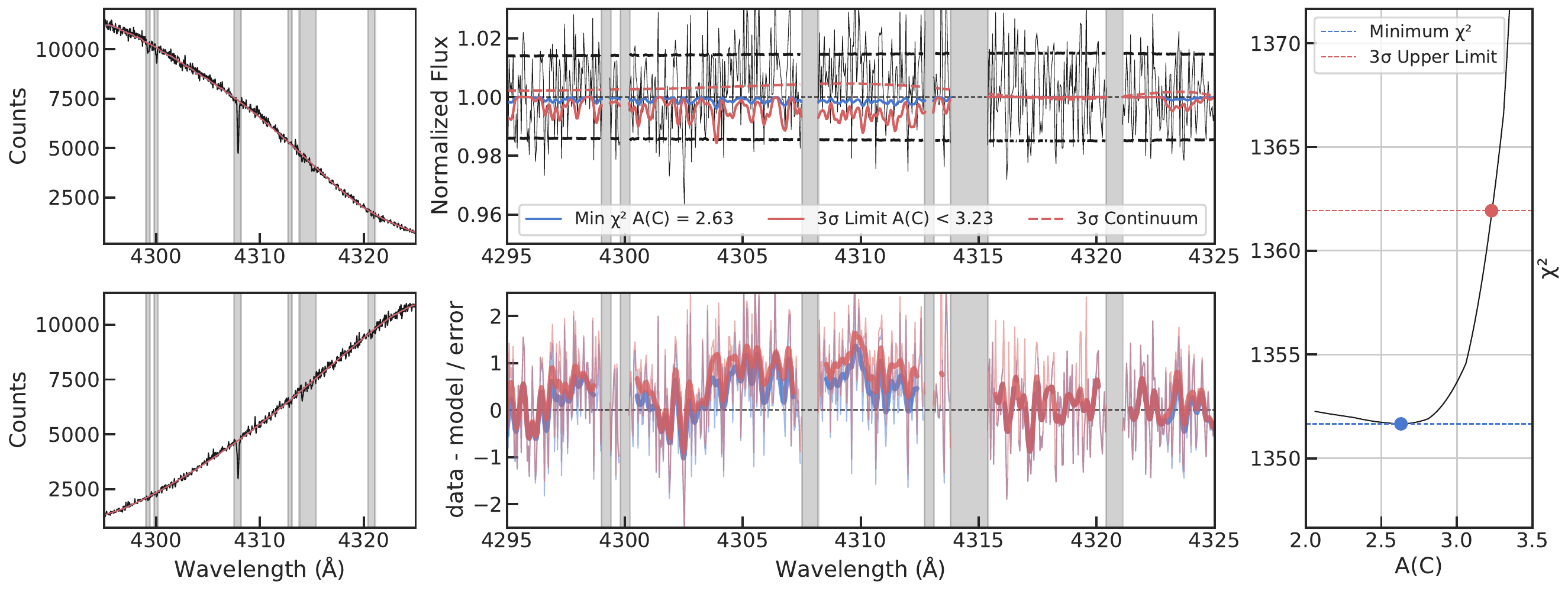}
    \caption{\textbf{Extended Data Figure 1.} Profile likelihood for CH upper limit. \textit{Left:} the two spectral orders being fit with a 7th degree polynomial, the data is fit well.
    A red model is plotted indicating the $3\sigma$ upper limit, masked regions shown in grey.
    \textit{Center top:} Stitched and normalized spectrum (black line, with $1\sigma$ pixel uncertainties shown as dashed black lines)
    compared to the best-fit (effectively no-carbon) spectrum (blue line) and the 3$\sigma$ upper limit (red line).
    The data and blue lines are normalized to 1, while the red $3\sigma$ model is normalized to the dashed red $3\sigma$ continuum line.
    The dashed red line is not exactly at 1, because the continuum is redetermined at every value of A(C), resulting in a more conservative upper limit when compared to a fixed continuum by about 0.2 dex.
    We apply an extra +0.2 dex correction to the final results, as the 3D model's stellar parameters are not identical to our adopted parameters (see text).
    \textit{Center bottom:} error-normalized residual for the best fit model (blue) and the $3\sigma$ upper limit (red). The per-pixel value is shown as a thin line, while the thick line is smoothed over 2 pixels.
    The red line is above the blue line where the CH features are.
    Note this is an approximation for visualization: the calculation is done on each order independently, not on the stitched spectrum.
    \textit{Right:} $\chi^2$ as a function of A(C).
    The blue point marks our minimum $\chi^2$ value.
    The $3\sigma$ upper limit, corresponding to 99.9\% confidence or $\Delta \chi^2 = 10.273$ for 1 degree of freedom, is marked as a red point.}
    \label{fig:chsynth}
\end{figure*}

\begin{figure*}
    \centering
    \includegraphics[width=\linewidth]{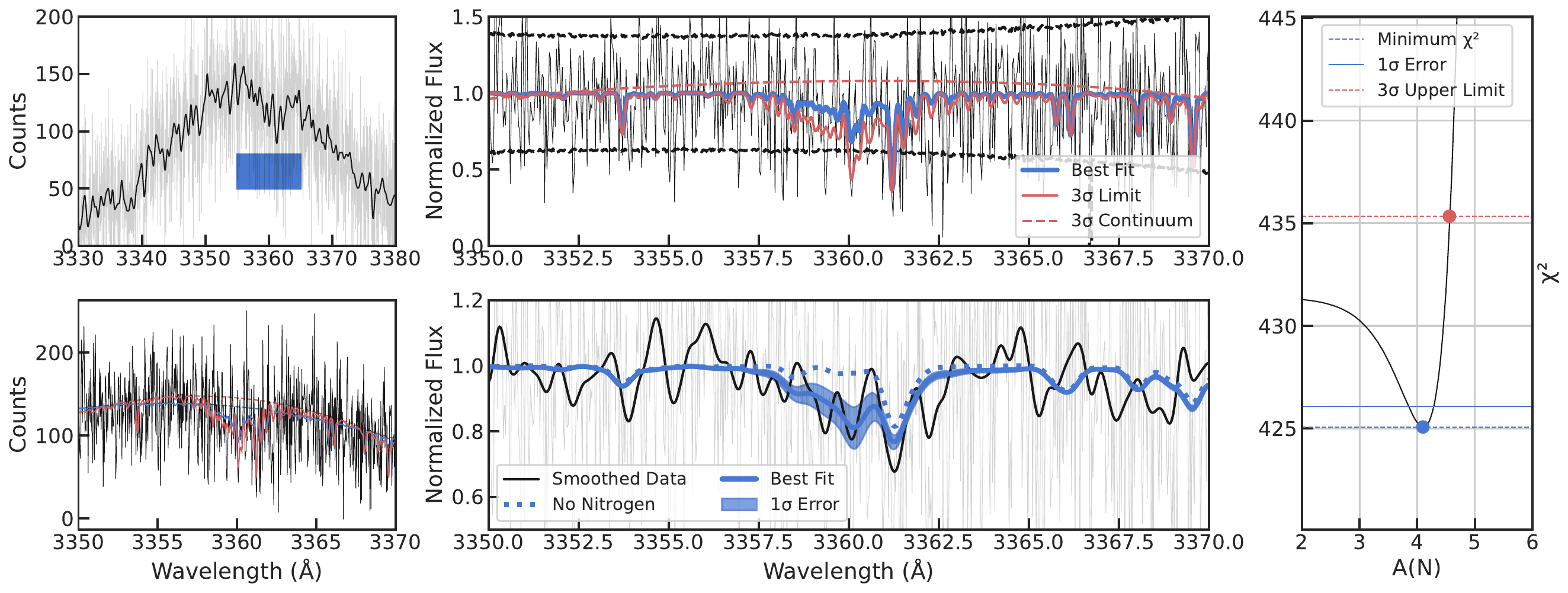}
    \caption{\textbf{Extended Data Figure 2.} Profile Likelihood for NH upper limit. \textit{Top Left:} the full spectral order containing the NH band. The black line is smoothed by a Gaussian with 5 pixel FWHM, and the blue box indicates a wavelength region where there is a clear deviation from the echelle order shape.
    \textit{Bottom Left:} the exact range being fit. The continuum (dashed lines) is modeled as a 2nd degree polynomial, which fits the data well.
    A blue model is plotted indicating the best fit, and a red model is plotted indicating the $3\sigma$ upper limit.
    \textit{Center top:} Normalized spectrum (black line, with $1\sigma$ pixel uncertainties shown as black dashed lines)
    compared to the best-fit model (blue line) and the 3$\sigma$ upper limit (red line).
    The data and blue lines are normalized to 1, while the red $3\sigma$ model is normalized to the dashed red $3\sigma$ continuum line.
    \textit{Center bottom:} Smoothed visualization of the data. The best-fit model and $1\sigma$ uncertainties are shown in solid blue line/shaded region, while the model with no nitrogen is shown as a dotted blue line.
    \textit{Right:} $\chi^2$ as a function of A(N).
    The blue point marks the best-fit nitrogen value, $\mbox{A(N)} = 4.10^{+0.18}_{-0.25}$.
    The $3\sigma$ upper limit is $\mbox{A(N)} < 4.56$ and marked as a red point.}
    \label{fig:nhsynth}
\end{figure*}

\vspace{1mm}
\noindent
{\bf Carbon and Nitrogen Upper Limits}
\\
\noindent
Special care was taken when deriving abundances or upper limits using molecular bands, namely the CH G band for carbon and the NH band at 3360{\AA} for nitrogen.
This is because for weak molecular bands, there is a well-known degeneracy between the continuum level and the strength of the molecular bands\cite{Caffau2024}, requiring careful treatment of normalization.
The typical analysis fixes the normalization and then places the upper limit by increasing the abundance until $\Delta \chi^2$ surpasses some set threshold\cite{Ji2020b, Lagae2023}.
Due to concerns about normalization accuracy, often only the strongest individual features are used\cite{Caffau2024}, but this neglects weaker features that should strengthen the upper limit.

Here, we use the profile likelihood\cite{Lampton1976,Feldman1998} to determine CH and NH upper limits that account for continuum normalization uncertainties.
Given a model spectrum with abundance $A(X)$ and nuisance continuum parameters $\mathcal{C}$, one can use a $\chi^2$ distribution model with one degree of freedom to describe the likelihood of $A(X)$, while accounting for $\mathcal{C}$\cite{Wilks1938}. 
In other words, we compute a grid of synthetic spectra at different values of $A(X)$, optimizing $\mathcal{C}$ at \textit{each} value of $A(X)$ and using the resulting $\chi^2$ contour to determine detections and upper limits.
For the purpose of determining upper limits, this is significantly more conservative than the standard approach: As $A(X)$ increases, the best-fit continuum level also increases, resulting in a higher $A(X)$ than if the continuum were previously fixed.

The carbon upper limit for \umpstar was determined using 3D LTE models of the CH G band.
The model atmosphere for this star was computed using the 3D radiation-hydrodynamics code \texttt{M3DIS}\cite{Eitner2024,Eitner2025}. We set the physical extent of the simulation domain to $10^{10}\,{\rm m}$ in the horizontal direction and $5\times10^{9}\,{\rm m}$ in the vertical direction. We then evolve the atmosphere with a grid resolution of $260\times260\times130$ points for $\sim 25000\,{\rm h}\approx 2.9\,{\rm yr}$ of stellar time to ensure that the final model is sufficiently relaxed. Due to the wide range of physical scales involved, it is very difficult and time-consuming to evolve 3D atmospheres of metal-poor giants towards a specific $\Teff$, as it is an output quantity in 3D rather than an input. Because the complexity further increases with increasing pressure scale height ($\sim T/g$) we chose to adopt a model that converged to $\Teff \approx 4660$\,K with $\logg=1.25$ and $\feh=-4.5$ for the 3D analysis, and later apply a correction for the difference. The post-processing synthesis of the G-band was done using an updated version of the \texttt{MULTI3D}\cite{Leenaarts2009,Eitner2024} code and uses the Masseron\cite{Masseron2014} CH linelist.
The CH strength depends on the oxygen abundance due to the C/O ratio, and we adopt a scaled-solar oxygen value in the analysis. We synthesized G-band spectra for A(C) ranging from $2.06$ to $4.31$ in increments of $0.25\,{\rm dex}$. Each synthetic spectrum was created by averaging the results from 10 snapshots of the same simulation, which were chosen to span at least one convective turnover time to properly account for the dynamic nature of the 3D atmosphere.

With the 3D model grid in place, the carbon upper limit was determined with the profile likelihood.
The signal-to-noise ratio in the G band region is 70/pixel or 100 per resolution element.
We fixed the spectral smoothing to 0.14{\AA} FWHM based on nearby detected lines and mask all absorption lines that are not due to carbon features.
The CH G band covers two different echelle orders on MIKE. Rather than normalize and stitch the two orders, we modeled the continuum of each order separately to avoid any spectral interpolation that would affect the noise properties. For each order, we adopt a seventh degree polynomial, which was selected by running our upper limit procedure for polynomial degrees between 3 and 13 and picking the lowest order model that minimized the Akaike information criterion\cite{Akaike1976}. The choice is purposefully conservative: lower polynomial degrees result in more stringent carbon upper limits by 0.1-0.2 dex, as they do not allow a slight continuum increase right above the carbon bands. Higher polynomial degrees $>11$ are sufficiently flexible to introduce an artificial $1-2\sigma$ detection by raising the continuum only above the CH bands while still matching the continuum outside of those regions.
We then adopted a $3\sigma$ (99.9\% confidence) upper limit, corresponding to $\Delta \chi^2 = 10.273$ for one degree of freedom. We also calculate a $5\sigma$ upper limit corresponding to $\Delta \chi^2 = 26.338$.

The results are shown in \ref{fig:chsynth}, giving a $3\sigma$ upper limit A(C) $< 3.23$.
The left panels show the actual data being fit, i.e. two spectral orders with the same spectral model but different continua and non-carbon regions being masked. The top middle panel shows a stitched version of the left panels, assuming the continuum fit from a model with almost no carbon. The dashed red line indicates the profile likelihood continuum for the $3\sigma$ limit, which differs from the minimum $\chi^2$ continuum because it is re-fit for each A(C) value.
The bottom-middle panel shows the error-normalized residual, where the statistical significance comes from integrating the difference between the red and blue curves. The right panel shows the difference in $\chi^2$ as a function of A(C), determining a 3D upper limit of A(C)$<3.23$, or $-0.65$ dex lower than our nominal 1D LTE abundance upper limit of A(C)$=3.88$.
To estimate the impact of using different stellar parameters, we also calculated the upper limits in 1D LTE at our fiducial $\Teff=4700$\,K and $\logg=1.10$ and the 3D atmosphere values $\Teff=4660$\,K and $\logg=1.25$. The upper limit differs by 0.20 dex, so we shift the 3D LTE abundance by the same value, resulting in A(C)$ < 3.43$. This is $-0.45$ dex lower than the 1D LTE upper limit.
We checked that an inaccurate wavelength calibration would result in ${<}$0.02 dex increase to the carbon upper limit.

It is worth noting that 1D atmospheres do not account for the adiabatic cooling that is present in 3D atmospheres and is responsible for decreasing the temperature in outer layers of the model atmosphere\cite{Asplund1999} and strengthening the carbon abundance limit\cite{Eitner2024,Caffau2024}. We thus also use a set of lower resolution models to estimate the differences in 3D.
We find that the changing temperature results in ${\lesssim}0.1\,$dex differences and changing $\logg$ results in ${\sim}0.2\,$dex differences, leading to a $+0.3\,$dex correction instead of the $+0.2\,$dex correction adopted above. Increasing the carbon abundance by $0.1-0.2\,$dex increases the total metallicity by $0.1\,$dex and would not change the conclusions of this paper.

During the first dredge up and after the red giant branch bump, mixing with the stellar interior brings up the products of the CNO cycle and increases the surface nitrogen abundance while decreasing the carbon abundance\cite{Gratton2000}.
Thus, the observed carbon abundance is lower than the natal carbon abundance, the observed nitrogen abundance is higher than the natal nitrogen abundance, and the total number of carbon and nitrogen atoms is conserved.
Stellar evolution models can be used to estimate the amount of carbon correction, and we use a precomputed correction grid by Placco et al.\cite{Placco2014}.
We find an evolutionary correction of $+0.56$ dex.
Since the physics of extra mixing and the carbon correction is somewhat uncertain and the star's \logg is near the red giant branch bump at this metallicity\cite{Fraser2022,Tayar2022}, we ran \code{MESA} models of the red giant branch as an independent check on the evolutionary correction, which obtained essentially identical results to the precomputed grid.
In summary, the total correction to the 1D LTE carbon abundance is $-0.45+0.56 = +0.11$ dex.
Thus, our final adopted carbon abundance limit is A(C) $< 3.99$ or $\mbox{[C/H]} < -4.52$, including both 3D and evolutionary effects.

The nitrogen constraint in such metal-poor stars comes from a very blue NH band at 3360{\AA}.
\ref{fig:nhsynth} shows this region of the spectrum, which is very noisy (signal-to-noise ratios of 3 per pixel, 4 per resolution element) as our star is a faint and cool red giant.
The smoothed version of the spectrum shows a possible NH detection visible at the strongest region at 3360{\AA}.
We again apply the profile likelihood to estimate the value of a detection or upper limit in 1D LTE with MOOG/ATLAS.
Working in the range $3350-3370${\AA}, we use a 2nd order polynomial to model the continuum, again chosen to minimize the Akaike information criterion.
The right panel of \ref{fig:nhsynth} shows the resulting $\chi^2$ profile. The minimum $\chi^2$ is achieved at $\mbox{A(N)} = 4.10^{+0.18}_{-0.25}$.
Unlike the case of CH, here we can visually detect a feature in the smoothed spectrum, and the polynomial order is also very low, leaving a possibility that this is a detection.
However, we conservatively adopt a $3\sigma$ (99.9\% confidence) upper limit of $\mbox{A(N)} < 4.56$ as the current nitrogen constraint. We expect the limit would decrease substantially in a 3D analysis similar to carbon. Additionally, the natal N abundance was lower depending on the number of C atoms that were converted to N during the MS lifetime. If we assume a detection at the A(C) upper limit, the carbon correction of $+0.56$ dex corresponds to a decrease in A(N) of $-0.09$ dex. We do not apply this downward correction to N, as it would be a smaller correction if the carbon abundance is below the upper limit.

\vspace{1mm}
\noindent
{\bf Kinematic Analysis and Magellanic Association}
\\
\noindent
The kinematic parameters of \umpstarsdss{} are estimated using the adopted spectro-photometric distance of $d_\mathrm{helio} = 26.1$~kpc and radial velocity $v_\mathrm{r} = 427.2~\kms$. 
All kinematic calculations are performed on $10^4$ Monte Carlo realizations of the  star's present-day phase-space position, with the median and half-difference between the 16th and 84th quantiles being adopted as the parameters with $1\sigma$ uncertainties. 
For the \textit{Gaia} DR3 proper motions, we utilize the full covariance information when generating these realizations. 
A right-handed Galactocentric frame is assumed with a solar position $\mathbf{x}_\odot = (-8.12, 0.00, 0.02)$~kpc, and solar velocity $\mathbf{v}_\odot = (12.9, 245.6, 7.8)$~$\mathrm{km\,s^{-1}}$ \cite{Reid2004,Drimmel2018,GravityCollaboration2019}. 

\ref{fig:lxlz} shows the kinematics of \umpstar as a large red star, other ultra-metal poor (UMP) stars and the LMC and SMC as colored points, and a shaded grey background of the entire SDSS-V halo sample\cite{Chandra2025}.
The top panel of \ref{fig:lxlz} shows the energy (assuming the static \code{MilkyWayPotential2022} potential from {\code{gala}}\cite{gala}) and $L_Z$ angular momentum component. Notably, \umpstar has total energy $> 0$, naively suggesting that it is unbound from the Milky Way. However, this ignores the influence of the Magellanic Clouds. 
The LMC (and SMC) are notable for their high orbital velocity, which manifests as a large angular momentum in the Galactocentric reference frame, particularly along the $L_\mathrm{X}$ coordinate. 
This extreme angular momentum signature is a powerful tool to search for stars that originate in the Clouds\cite{Chandra2023b}.
Thus the bottom panel of \ref{fig:lxlz} shows the space of Galactocentric angular momentum in the $L_\mathrm{Z}-L_\mathrm{X}$ space. 
\umpstar has angular momenta similar to the LMC and SMC, and falls securely in the region of angular momenta occupied by Magellanic Debris.
\umpstar has a negligible angular momentum in the unseen dimension, with $L_\mathrm{Y} \approx 0.9\,\kms$\,kpc. 
Therefore, \umpstarsdss{} has an orbital trajectory similar to the Magellanic Clouds. 

\begin{figure}%[h]
    \centering
    \includegraphics[width=\linewidth]{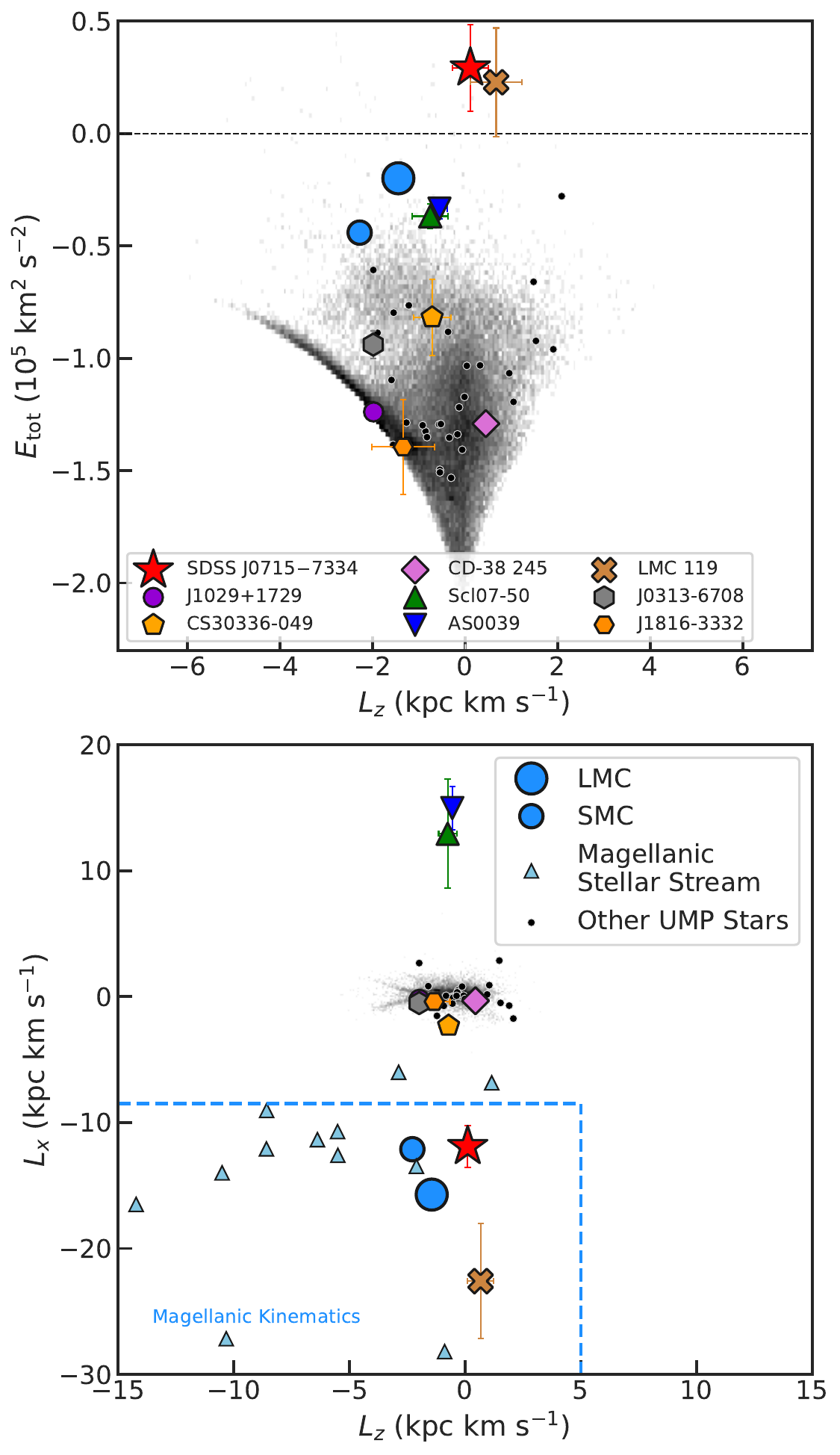}
    \caption{\textbf{Extended Data Figure 3.} Energy and angular momentum in a static potential.
    \textit{Top:} Specific energy and angular momentum. \umpstar is shown as a large red star. Black points show a literature sample (see Literature Data Sample section in text). Large blue circles indicate the LMC and SMC. 
    Colored points highlight eight notable metal-poor stars with 68\% confidence uncertainties.
    The same stars are shown in the Main Text in Figure~\ref{fig:cfe}.
    The shaded grey background is metal-poor stars from the SDSS-V halo program, computed in a static Milky Way gravitational potential (see Methods). \umpstar joins LMC-119\cite{Chiti2024} as originating from the LMC.
    \textit{Bottom:}
    Galactocentric specific angular momentum in the $L_\mathrm{Z}-L_\mathrm{x}$ plane, in which stars associated with the Magellanic Clouds have a distinctive signature\cite{Chandra2023b}. 
    All halo stars from SDSS-V are shown, along with known ultra-metal-poor stars from the literature with 68\% confidence uncertainties.
    Magellanic Stellar Stream members proposed by Chandra et al.\cite{Chandra2023b} are shown, along with the selection box used to identify Magellanic Debris. 
    \umpstar has kinematics that strongly associate it with the Clouds.}
    \label{fig:lxlz}
\end{figure}

To further investigate the past orbit of \umpstar, we integrate back its phase-space position in a time-varying potential that includes the gravitational influence of the LMC \cite{Patel2020, Chandra2023b}. 
This calculation is implemented in \texttt{agama}\cite{Vasiliev2019}. 
Specifically, we adopt the \texttt{MilkyWayPotential2022} from \texttt{gala}\cite{gala}, which has a circular velocity at the solar position $v_\mathrm{c}(R_\odot) = 229~\textrm{km}~\textrm{s}^{-1}$ \cite{Eilers2019}. 
The LMC is modeled as a rigid $1.8 \times 10^{11}\,M_\odot$ NFW profile that matches current observational constraints\cite{Vasiliev2021a}. 
The LMC is integrated back from its current position taking into account dynamical friction, which defines the time-varying potential\cite{VanDerMarel2002, Pietrzynski2019, Luri2021}. 
The orbit of \umpstar is subsequently integrated in this time-varying potential, again for $10^4$ Monte Carlo realizations of the present-day phase-space position. 
A typical (median) orbit is shown in Figure~\ref{fig:kinematics}, which also shows the UMP stars \caffaustar{} and LMC-119 integrated in the same potential for comparison. 
\caffaustar{} remains confined to the MW disk, whereas LMC-119 is firmly bound to the LMC as expected.

\begin{figure}
    \centering
    \includegraphics[width=\linewidth]{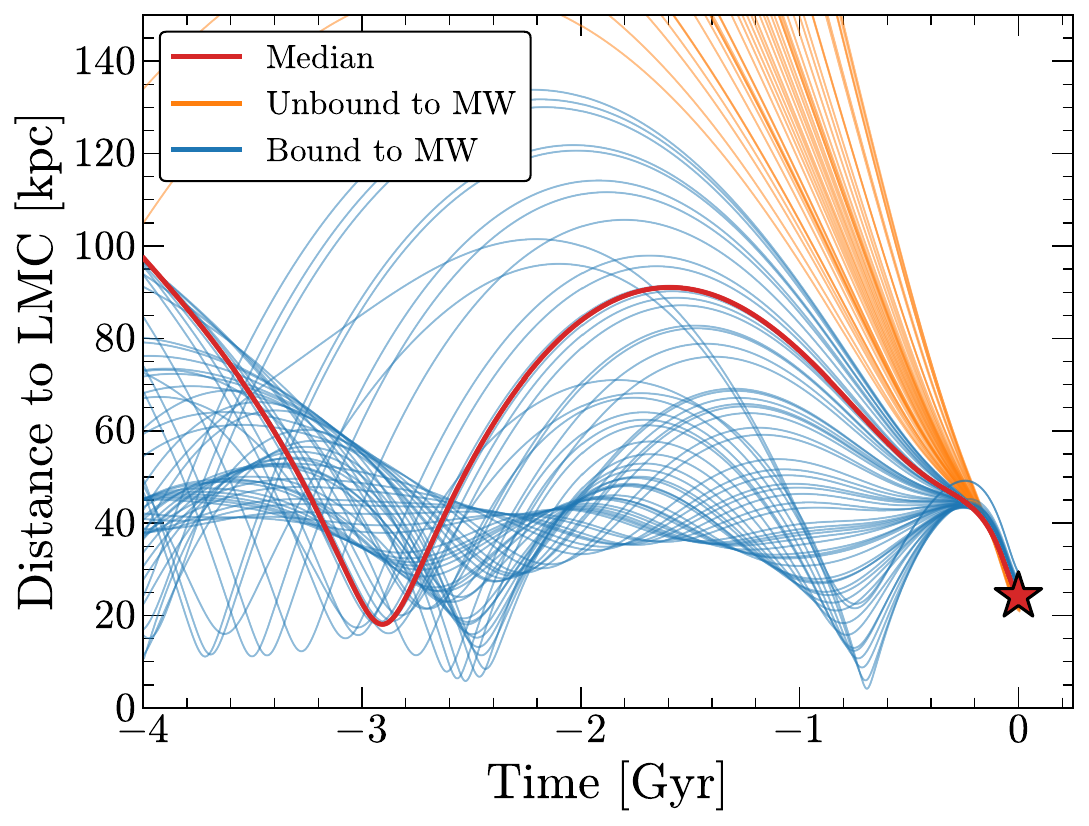}
    \caption{\textbf{Extended Data Figure 4.} 
    Distance to LMC over time for different orbit integration samples.
    The thick red line shows the median orbit. The same orbit is shown in the Main Text in Figure~\ref{fig:kinematics}. The other lines are color-coded by the \textit{future} fate of \umpstar: orange lines show orbits that will be unbound to the Milky Way, while blue lines show orbits that will remain bound to the Milky Way.}
    \label{fig:lmcdistsamples}
\end{figure}

A random subset of orbits is shown in \ref{fig:lmcdistsamples}, color-coded by the future fate of \umpstar.
About 60\% of orbits have a pericenter with respect to the LMC of 5-50kpc, and apocenters within the LMC virial radius (${\sim}120$kpc). In other words, they clearly originate from the LMC. Indeed, compared to the Magellanic Stellar Stream members from Chandra et al.\cite{Chandra2023b}, \umpstarsdss{} has one of the most confident kinematic associations with the LMC.
However going forward, these orbits do not remain bound to the LMC. 
They are highly eccentric orbits with apocenters with respect to the Milky Way ranging from $150-250$~kpc, while the LMC continues to sink deeply into the Milky Way potential due to dynamical friction. Thus, these solutions (shown in blue in \ref{fig:lmcdistsamples}) correspond to \umpstar being a LMC halo star that comes in with the LMC system and subsequently remains bound to the Milky Way.

The other 40\% of orbits are not previously bound to the LMC, i.e. they do not have a close pericenter with the LMC in the last 4 Gyr. 
However, these same orbits are completely unbound and will leave the combined MW-LMC system in the future (orange lines). These solutions correspond to stars originating from the cosmic web with such high peculiar velocities that they pass straight through the Milky Way, but coincidentally they line up with the Magellanic Clouds' current orbital configuration. Such solutions are highly implausible compared to the case that \umpstar originates from the LMC system.

The detailed orbit within the LMC has substantial uncertainties (\ref{fig:lmcdistsamples}). About half the orbits have more eccentric orbits and tighter pericenters $<20$kpc, while the other half have more circular orbits and larger pericenters. There are additional modeling uncertainties, such as the LMC's mass and halo shape, the amount of dynamical friction, and the interaction with the SMC.
These uncertainties should not alter the association of \umpstar with the LMC, but would affect whether \umpstar is best interpreted an in situ LMC halo star or part of the LMC's accreted halo.

\vspace{1mm}
\noindent
{\bf Comparison to Limberg et al. 2025}

Recently, Limberg et al.\cite{Limberg2025} made an independent identification of the same star using \textit{Gaia} XP spectra\cite{Yao2024}.
Their followup spectra also used Magellan/MIKE, but only 40 minutes of data compared to our 225 minutes. They obtained stellar parameters $\Teff=4596 \pm 65$\,K, $\logg=0.88 \pm 0.15$ cgs, $\xi_t=2.27\pm0.10\kms$, and $\feh = -4.82 \pm 0.25$ from a 1D LTE analysis.
The metallicity is 0.3 dex lower than our 1D LTE analysis.
Their photometrically determined temperature is 100\,K smaller than our fiducial value which is at the coolest end of what we infer, and they adopted the Yonsei-Yale isochrones\cite{Yi2001,Demarque2004} to determine $\logg$ which accounts for that difference.
Due to the lower signal-to-noise ratio of that spectrum, they adopt an empirical $\logg-\xi_t$ relation\cite{Ji2023} that results in a much higher $\xi_t$, but this higher $\xi_t$ is inconsistent with our measured Fe\,I line strengths.
If we restrict only to the strongest, relatively saturated lines measured in our spectrum and adopt their stellar parameters, we can reproduce their conclusions. About half of the 0.3 dex difference in [Fe/H] is attributed to their lower temperature and the other half to their higher microturbulence. 
Due to the low signal-to-noise ratio of the spectrum, they could only put a relatively high upper limit $\mbox{[C/Fe]}  < +0.5$ ($\mbox{[C/Fe]}_{\rm corr} < +1.2$ after evolutionary corrections) and measured element abundances from fewer lines resulting in abundances consistent with ours within the uncertainties. 

The higher S/N ratios, improved upper limit analysis, and additional NLTE and 3D analyses in this work are necessary for a stringent total metallicity constraint and comparison to theoretical yield models that were not possible by Limberg et al. 
In particular, we can put a $D_{\rm trans}$ constraint that is clearly below the critical threshold. They derived $\mbox{[C/H]} < -3.64$, which results in $D_{\rm trans} = -3.5 \pm 0.1$, right at the threshold $D_{\rm trans} = -3.5 \pm 0.2$. We derive $D_{\rm trans} = -4.2^{+0.3}_{-0.2}$, clearly below the threshold. We are also able to perform a meaningful comparison to supernova yields with the more accurate NLTE chemical abundances and better abundance precision.

In principle our results could be improved by combining both datasets. We tested combining their data with ours and rerunning the analysis, finding a negligible change in results. To illustrate, we focus on the CH upper limit that is most sensitive to improved signal-to-noise ratios. In the G band region, the median signal-to-noise of the coadded spectrum would have increase from 70/pixel to 75/pixel and naively result in a better upper limit. However, we find no change to the upper limit when including the new data. This is because we optimize over uncertainties in continuum placement for each new order instead of rebinning and coadding the data (to avoid introducing pixel correlations that affect the $\chi^2$ analysis), and the uncertainties in continuum placement remove the benefit of the higher total S/N ratio. We thus use only our higher quality spectrum in our analysis.

Limberg et al. also perform a kinematic analysis. The radial velocity and proper motion data are identical to ours, and in particular there is no radial velocity variation between the two MIKE observations. They adopt a slightly larger distance due to their lower \logg value. 
Although Limberg et al. also find evidence for an association with the LMC, with over 50\% of their orbit calculations having a long-term pericenter below 60 kpc\cite{Chandra2023b}, we here have also shown that the orbits not bound to the LMC are unbound from the Milky Way (Figure~\ref{fig:lmcdistsamples}). We thus claim a much stronger likelihood of association with the LMC than Limberg et al.

\begin{figure}
    \centering
    \includegraphics[width=\linewidth]{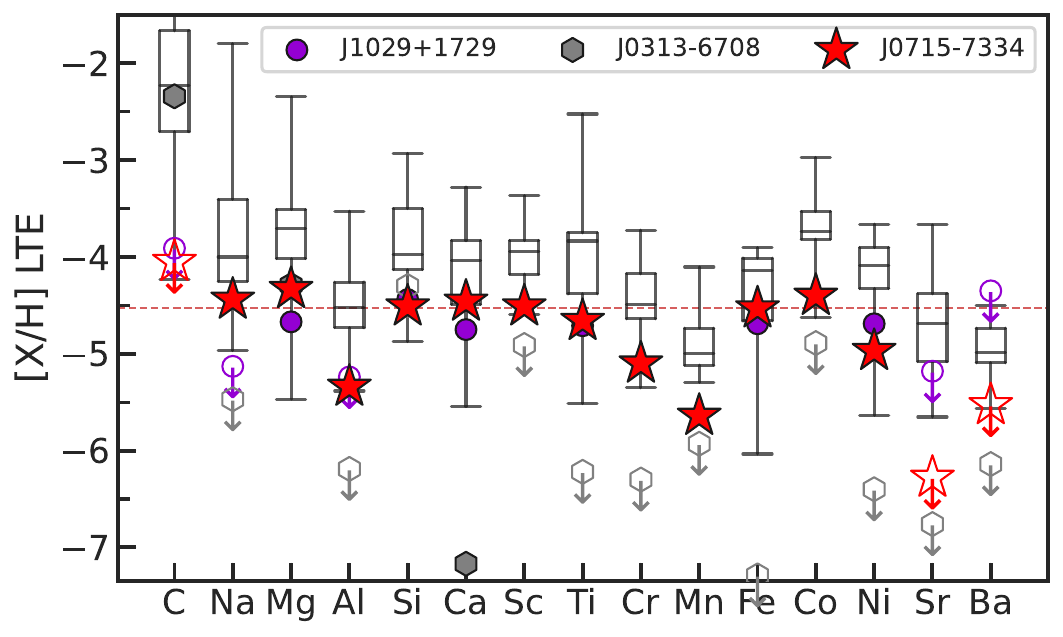}
    \caption{\textbf{Extended Data Figure 5.} 1D LTE abundances of \umpstar for different elements, compared to literature stars. 
    Grey boxplots indicate the minimum, maximum, median, and 25-75 percentile [X/H] range for the literature stars in 1D LTE, where carbon has evolutionary corrections.
    Colored points highlight the 1D LTE analyses of {\kellerstar}\cite{Keller2014} and {\caffaustar}\cite{Caffau2024}
    A horizontal red line is drawn at the [Fe/H] value for \umpstar.
    }
    \label{fig:boxplot}
\end{figure}

\vspace{1mm}
\noindent
{\bf Literature Data Sample}
\\
\noindent
We identified 38 stars\cite{Cayrel2004,Collet2006,Norris2007,Frebel2008,Lai2008,Cohen2008,Caffau2011,Yong2013,Keller2014,Roederer2014,Hansen2015,Placco2015,Howes2015,Bonifacio2015,Simon2015,Frebel2015,Li2015,Melendez2016,Placco2016,Caffau2016,Bonifacio2018,Starkenburg2018,Aguado2019,Frebel2019,Nordlander2019,GonzalezHernandez2020,Placco2024,Chiti2024,Skuladottir2024} with $\mbox{[Fe/H]}_{\text{1D, LTE}} \lesssim -4$ in the literature, shown in Figures~\ref{fig:cfe} and \ref{fig:kinematics}.
The stars were primarily identified in the JINABase literature compilation\cite{Abohalima2018} and supplemented using the SAGA literature compilation\cite{Suda2008} and other recent references.
We excluded stars that only had low-resolution spectroscopy and thus only Fe and C abundances, though many such stars are known\cite{Aguado2017,AllendePrieto2023}.

We put a special focus on eight stars due to their extremely low total metallicities, extreme abundances, and/or orbits:
the previous record holder for the most metal-poor star J1029$+$1729\cite{Caffau2011,Caffau2012,Lagae2023,Caffau2024}, the first nearly ultra-metal-poor star in the inner galaxy J1816$-$3332\cite{Howes2015}, the most iron-poor star known J0313$-$6708\cite{Keller2014,Nordlander2017}, the previous LMC metal-poor record holder LMC-119\cite{Chiti2024}, the first most metal-poor star known {\cdstar}\cite{Bessell1984,Cayrel2004,Mittal2025}, two ultra-metal poor stars in the Sculptor dSph Scl07-50\cite{Simon2015} and AS0039\cite{Skuladottir2021,Skuladottir2024}, and an ultra-metal-poor halo star CS30336-049\cite{Lai2008}.
The stars J1029$+$1729\cite{Lagae2023} and J0313$-$6708\cite{Nordlander2017} have full 3D NLTE analyses where possible, and 3D LTE for CH and NH.
The stars AS0039\cite{Skuladottir2024} and \cdstar\cite{Mittal2025} have full 1D NLTE analyses.
For the other four stars, we determined 1D NLTE corrections to the Fe abundances\cite{Bergemann2012b} and carbon evolutionary corrections\cite{Placco2014} to place them on Figure~\ref{fig:cfe}.
A star of note that is not included is Pristine\_221\cite{Starkenburg2018}, which has the potential to be similarly metal-poor with $\feh_{\rm LTE} = -4.8$. However, this star has a very loose carbon upper limit $\mbox{[C/H]} < -2.5$\cite{Lardo2021}, so it may still have a high overall metallicity.

The detailed chemical abundances of the whole sample in 1D LTE are shown in \ref{fig:boxplot}, highlighting \umpstar, \caffaustar, and \kellerstar.
We use 1D LTE here as NLTE and 3D analyses are not available for the vast majority of stars, but we apply evolutionary corrections to carbon\cite{Placco2014}.
The background sample is shown as a boxplot for each element.
\umpstar has one of the lowest carbon abundances of any star known, which along with its very low iron abundance is what makes it the most metal-poor star currently known. It also has the second-lowest Sr and Ba abundances known.
\kellerstar has lower abundances of almost every element including the lowest Fe abundance known, but it has such high C (and O\cite{Bessell2015}) abundances that its total metallicity is quite high $\logzzsun = -2.7$.
\caffaustar is a relatively faint and ultra-metal poor dwarf star, so fewer elements are available but enough have been measured to determine a strong metallicity upper limit and abundance pattern\cite{Caffau2012,Lagae2023,Caffau2024}. Especially notable is that it has a very low Na and Mg abundance, which has been attributed to either an explosion of a relatively low-mass star ($10-20 M_\odot$)\cite{Lagae2023} or surface pollution from the interstellar medium\cite{Caffau2024}.
We emphasize that not all core-collapse supernovae must produce high [Mg/Fe] ratios; the Mg production correlates strongly with a star's initial mass and can be low for the lowest mass core collapse supernovae\cite{Jeena2024}.

We also computed kinematics for all 38 stars consistently with \umpstar.
Radial velocities were obtained from the literature references.
Distances were derived using photometry and parallax from the \code{isochrones} package, with the same extinction, metallicity, and distance priors as before.
For the three stars known to be in Sculptor and the LMC, we use the distance of the host galaxy as a prior.
One star (SDSS~J174259.67+253135.8) has insufficient information for a good distance so is not included in the kinematic compilation.
Overall, we draw the same qualitative conclusions as the kinematic study by Sestito et al.\cite{Sestito2019}, including that a vast majority of stars with low $|z_{\rm max}|$ are prograde, and that \caffaustar is on a thick disk orbit consistent with forming in the proto-Milky Way or in an extremely early accretion event\cite{Mardini2022}.

\vspace{1mm}
\noindent
{\bf Total Metallicity Calculation}
\\
\noindent
The total metallicity $Z$ of a star is given by\cite{Lagae2023}:
\begin{equation*}
    Z = \frac{1 - Y}{1 + X/Z}
\end{equation*}
where we assume a primordial helium mass fraction $Y = 0.2477$\cite{Peimbert2007} and the metals-to-hydrogen mass ratio is
\begin{equation*}
    \frac{Z}{X} = \sum_i \left[10^{A(i)-12} \mu(i)\right]
\end{equation*}
where $A(i)$ is the abundance of element $i$ and $\mu(i)$ is the atomic weight of element $i$.

As not all elements are measured in stars, one must extrapolate the missing elements to determine the metallicity.
In the main text we report a metallicity calculated the simplest way (Method 1), which is to assume each missing element scales with overall iron abundance such that $\mbox{[X/Fe]}=0$ for all elements X, i.e. fill them in with solar ratios.
We adopt the recent Lodders et al.\cite{Lodders2025} solar abundance pattern for the solar ratios, which has total metallicity $Z_\odot=0.016$.
This provides $Z < 7.85 \times 10^{-7}$ or $\logzzsun < -4.31$.
A possibly more accurate way is to correlate the missing elements with another element that more closely traces the nucleosynthetic pathways that produce a particular element, e.g. [X/Mg]$=0$ for elements correlated with magnesium.
As a first estimate, we associate C, N, and O with the carbon abundance (light elements); F through Ca with the magnesium abundance (alpha-elements); and Sc and up with the Fe abundance (iron peak and neutron capture elements).
This assumption (Method 2, motivated by Ref. \cite{Caffau2024}) leads to a slightly lower value $Z < 7.61 \times 10^{-7}$ or $\logzzsun < -4.32$.
To estimate metallicities for many stars, if only C, Mg, and Fe are measured, one can derive an equation for the total metallicity:
\begin{equation*}
    Z = 31.22 \times 10^{\text{A(C)}-12} + 118.93 \times 10^{\text{A(Mg)}-12} + 45.69 \times 10^{\text{A(Fe)}-12}
\end{equation*}
This equation, assuming $\mbox{[Mg/Fe]}=+0.4$, is used to calculate the blue contours in Figure~\ref{fig:cfe}.

As the most abundant metal that is almost always unmeasured, the assumption for the oxygen scaling is the most important systematic effect.
If instead oxygen scales as an alpha-element with magnesium instead of carbon (Method 3), we obtain $Z < 10.5 \times 10^{-7}$ or $\logzzsun < -4.18$.
If we hold all missing elements at solar composition then fix $\mbox{[O/Fe]}=+0.6$ (Method 4, Following Ref. \cite{Lagae2023}), this gives $Z < 18.0 \times 10^{-7}$ or $\logzzsun < -3.95$.
Finally if we adopt $\mbox{[C/O]}=-0.6$ similar to most other metal-poor stars\cite{Amarsi2019}, this gives $Z < 21.9 \times 10^{-7}$ or $\logzzsun < -3.86$ (Method 5).
Under almost all of these assumptions, \umpstar remains the most metal-poor star known, with \caffaustar being the second-most metal-poor star.
We calculate that for Methods 1-5, \caffaustar has metallicities $\logzzsun < -4.23, -4.29, -4.34, -3.86$, and $-3.61$ respectively. 
Note that we derive a lower metallicity upper limit for \caffaustar than Caffau et al.\cite{Caffau2024} when using the same assumptions (Method 2) because we also adjust N and O when applying the 3D correction to carbon, while they only change carbon.
Under these varying assumptions, \caffaustar has a total metallicity upper limit consistently 0.05-0.25 dex more metal-rich than \umpstar's metallicity upper limit. The exception is if O scales with Mg, due to the higher Mg abundance in \umpstar.
Redoing the calculations assuming a solar pattern with lower oxygen abundance\cite{Asplund2021} gives 0.06 dex lower total metallicities throughout. Either way, these two stars clearly stand out from other known stars as having $\logzzsun < -4$ under all assumptions, thus robustly requiring that they form under the influence of dust cooling\cite{Schneider2012}.

There are currently some discrepancies in the literature on the chemical abundances of \caffaustar. Here, we adopt the 3D NLTE abundances (or 3D LTE when not available) by Lagae et al.\cite{Lagae2023} for all elements except carbon, where we adopt the 3D LTE abundances from a recent analysis by Caffau et al.\cite{Caffau2024}. This choice is made as we consider the 3D NLTE analysis to include the maximum amount of relevant physics for analyzing metal-poor stars, but Caffau et al. analyze a spectrum with about 4 times more data which should produce the strongest upper limit on carbon.
Focusing on the 3D and/or NLTE abundances of key elements, Lagae et al. find $\mbox{A(C)} < 4.86$, $\mbox{A(Mg)} = 3.18$, and $\mbox{A(Fe)} = 3.28$. Caffau et al. find $\mbox{A(C)} < 4.11$, $\mbox{A(Mg)} = 3.00$, and $\mbox{A(Fe)} = 2.91$. If we adopt all of the Caffau et al. abundances, we find the two stars \umpstar and \caffaustar have very similar total metallicities primarily due to a much smaller NLTE correction to Fe, and which star is more metal poor depends entirely on the assumption of the abundance of missing elements. This is evident if we calculate a lower limit on the total metallicity by setting all unmeasured elements to 0 abundance: both \umpstar and \caffaustar have $Z > 1.84 \times 10^{-7}$ or $\logzzsun > -4.94$. If the Caffau et al. abundances are more accurate, then much deeper observations that constrain more elements in both stars, especially oxygen, will be needed to discern which star has a more metal-poor upper limit.
We note that it is unlikely in the near future to obtain such constraints on \caffaustar as there are no detections in 29 hours of UVES data, but as a cool and fairly bright red giant there is a chance that OH and NH could be detected in \umpstar.

\vspace{1mm}
\noindent
{\bf Population~III Yield Fits}
\\
\noindent
We use the metal-free core collapse supernova yield models by Heger \& Woosley\cite{Heger2010} to fit the chemical abundance pattern of \umpstar.
Due to uncertainties in the underlying yield models, we follow the recommendation to ignore our Cr measurement and Zn upper limit, and we treat the model prediction for Sc as a lower limit (i.e., our measured Sc is an upper limit), which is indicated by different size points shown in Fig~\ref{fig:popiii}.
We did include the non-constraining upper limits on O and K in the fit, but these are not plotted in Figure~\ref{fig:popiii} for clarity.
The best-fit models are found by minimizing the error-normalized total absolute error (TAE)\cite{Ji2024}:
\begin{equation*}
    \text{TAE} = \sum_X \left| \frac{\text{[X/H]}_{\rm obs} - \text{[X/H]}_{\rm model}}{\sigma_X}\right|
\end{equation*}
We use the TAE instead of $\chi^2$ because we expect the fits to be dominated by systematic uncertainties, and the TAE penalizes outliers less than $\chi^2$.
This is then an optimization over 4 parameters: the progenitor mass $M$, the final kinetic explosion energy $E$, the mixing parameter $\xi$, and the dilution factor $D$. In practice, we find the best-fit $D$ for all 16800 models through a brute force search initialized at the analytic minimum $\chi^2$ solution. We reject all models that are inconsistent with our upper limits, then take the best-fit result as the model with the smallest TAE.

We adopt a weight for each of the 16800 models assuming the data are distributed according to a Laplace distribution, which is appropriate for minimizing the TAE. The per-species uncertainty $\sigma_X$ is turned into a Laplace scale $b_X=\sigma_X/\sqrt{2}$. Assuming the likelihoods satisfy Wilks' Theorem\cite{Wilks1938}, we calculate twice the negative log likelihood and assign a weight $w \propto \exp\left[-\sqrt{2}\, \text{TAE} \right]$.
We use these weights to estimate the best-fit progenitor mass and explosion energy for \umpstar as a weighted sum and standard deviation, finding $M = 27.0 \pm 3.9 M_\odot$ and $E = (6.0 \pm 2.6) \times 10^{51}$\,erg, although the energy uncertainty is strongly affected by discreteness effects.
We also use these weights to construct the weighted histogram in the bottom left-hand panel of Figure~\ref{fig:popiii} and determine the opacity and size of lines and points in the other two panels. As each plotted line has the same opacity, the thick black regions in Figure~\ref{fig:popiii} indicate areas where many models overlap.

\vspace{1mm}
\noindent
{\bf Critical Metallicity Calculations (Whole Section Is New)}
\\
\noindent
The critical metallicity for atomic fine structure cooling is driven primarily by the carbon and oxygen line cooling\cite{Bromm2003}. It can be described as\cite{Frebel2007}:
\begin{equation*}
    D_{\rm trans}=\log_{10}(10^{\mbox{[C/H]}} + 0.3 \times 10^{\mbox{[O/H]}}) > -3.5 \pm 0.2
\end{equation*}
which is similar to $\logzzsun > -3.5$. This threshold assumes that nearby Lyman Werner radiation is sufficient to dissociate all molecular hydrogen in the collapsing gas cloud, which could otherwise somewhat lower the critical metallicity threshold by providing some additional cooling\cite{Smith2024}.

As none of the most metal-poor stars have oxygen measurements, computing the $D_{\rm trans}$ criterion requires an assumption for the [O/H] abundance. In this metal-poor regime, it is physically most appropriate to choose [O/H] based on the [C/O] ratio (instead of e.g. an [O/Fe] ratio), because C and O have tighter correlations both observationally\cite{Suda2008,Amarsi2019} and theoretically from core-collapse supernova nucleosynthesis calculations\cite{Heger2010,Ji2024}. For Figure~\ref{fig:cfe}, the horizontal dark red line indicates the conversion from 
$D_{\rm trans}=-3.5$ to a critical $\mbox{[C/H]}=-3.84$ by adopting the typical value $\mbox{[C/O]}=-0.6$ from the [C/O] plateau in metal-poor stars\cite{Amarsi2019}. The narrow shaded uncertainty region ranges from $-4.10 < \mbox{[C/H]} < -3.61$, corresponding to a fairly extreme range of values $-1 < \mbox{[C/O]} < 0$ that spans most data and nucleosynthesis models\cite{Suda2008,Heger2010}. Higher [C/O] ratios are possible (e.g. the carbon-enhanced metal-poor stars) but would not affect the total metallicity. The widest uncertainty region adds $\pm 0.2$ dex on top of this range, due to theoretical uncertainties in the temperature and density of gas assumed to calculate $D_{\rm crit}$\cite{Bromm2003}.
Even in the extreme case $\mbox{[C/O]} = -1$ and $D_{\rm trans}=-3.7$, \umpstar is clearly below the fine structure cooling threshold. It is only the second star to be below this threshold, along with \caffaustar\cite{Caffau2024}. 

The critical metallicity for dust thermal cooling is driven primarily by silicate grains (e.g. enstatite MgSiO$_3$ and forsterite Mg$_2$SiO$_4$) or amorphous carbon grains\cite{Bianchi2007,Schneider2012,Ji2014,Chiaki2015,Chiaki2017}. Amorphous carbon is not important for \umpstar due to its low carbon abundance\cite{Chiaki2017}.
It is possible to calculate a critical abundance threshold for dust cooling, though it depends on the grain species being considered, the grain size distribution which can be characterized by characteristic size $a_{\rm cool} = \langle a^3 \rangle / \langle a^2 \rangle$, and the dust depletion factor $0 \leq f_{\rm dep} \leq 1$\cite{Schneider2012,Ji2014,Chiaki2017}.

Chiaki et al.\cite{Chiaki2017} provide an estimate for the critical Mg abundance assuming silicates can be characterized by enstatite, whose production is limited by the key element Mg:
\begin{equation*}
    \textrm{[Mg/H]}_{\rm cr} = -4.70 + \log \left(\frac{a_{\rm cool}/f_{dep,\text{MgSiO}_3,\text{Mg}}}{0.1 \mu \text{m}}\right)
\end{equation*}
Given an [Mg/H] measurement, we can put a lower limit on the depletion factor of Mg by rearranging:
\begin{equation*}
    f_{dep,\text{MgSiO}_3,\text{Mg}} > 10^{-\mbox{[Mg/H]}-4.70} \left(\frac{a_{\rm cool}}{0.1\mu\text{m}}\right)
\end{equation*}
For \umpstar with $\mbox{[Mg/H]} = -4.07$ and assuming a typical grain size of 0.01-0.1 $\mu\text{m}$\cite{Nozawa2007,Schneider2012}, this gives $f_{dep,\text{MgSiO}_3,Mg} \gtrsim 0.02-0.2$.
Doing the same calculation for forsterite (key element Si) gives
\begin{equation*}
    f_{dep,\text{Mg}_2\text{SiO}_4,\text{Si}} > 10^{-\mbox{[Si/H]}-4.75} \left(\frac{a_{\rm cool}}{0.1\mu\text{m}}\right)
\end{equation*}
or for \umpstar with $\mbox{[Si/H]} = -4.27$ gives $f_{dep,\text{Mg}_2\text{SiO}_4,\text{Si}} \gtrsim 0.03-0.3$.
We conservatively estimate that the minimum depletion factor of these elements is 1\%.
This potentially places constraints on the physics of early dust production: because the explosion energy we infer for \umpstar's supernova progenitor is quite high ($E_{\rm expl} \gtrsim 5 \times 10^{51}$ erg), we expect the supernova reverse shock would be especially effective at destroying dust grains\cite{Bianchi2007}, leading to low dust depletion that can be well below 1\% after a strong shock\cite{Schneider2012}. However, dust grain growth during protostellar collapse can increase the depletion factor back up to ${\gtrsim}1\%$\cite{Chiaki2015}.
A similar conclusion can be reached using \caffaustar\cite{Schneider2012b,Chiaki2014}, and it is somewhat remarkable that, despite the very different formation environments and supernova progenitors (mass, explosion energy), the conclusions about dust properties can be similar.

\begin{addendum}
  
\item [Acknowledgements] 

We acknowledge The College at University of Chicago for their support of undergraduate research that led to the identification of this star and supporting its analysis.
This paper includes data gathered with the 6.5 meter Magellan Telescopes located at Las Campanas Observatory, Chile. We thank the staff at Las Campanas Observatory for their support making the observations possible.
APJ thanks Alex Drlica-Wagner, Harley Katz, Jenny Greene, and Diogo Souto for useful discussions; 
and Ian Roederer, Ian Thompson, and Steve Shectman for a comparison spectrum of \cdstar.
We thank the reviewers for comments that significantly improved this article.
We acknowledge support from the National Science Foundation under awards 
AST-2206264 (APJ, SMT, ZZ, PNT),
AST-2338645 (KCS),
and DGE2139841 (WC).
APJ acknowledges the Alfred P. Sloan Research Fellowship and the University of Chicago's Research Computing Center.
MB is supported through the Lise Meitner grant from the Max Planck Society and through the European Research Council (ERC) under the European Unions Horizon 2020 research and innovation programme (Grant agreement No. 949173). 
MH and JAJ acknowledge support from NASA grant 80NSSC24K0637.
CFPL acknowledges funding from the European Research Council (ERC) under the European Union’s Horizon 2020 research and innovation programme (grant agreement No. 852839) and the Agence Nationale de la Recherche (ANR project ANR-24-CPJ1-0160-01).
WC acknowledges support from a Gruber Science Fellowship at Yale University.
JGF-T acknowledges the support provided by ANID Fondecyt Regular No. 1260371, ANID Fondecyt Postdoc No. 3230001 (Sponsoring researcher), the Joint Committee ESO-Government of Chile under the agreement 2023 ORP 062/2023 and the support of the Doctoral Program in Artificial Intelligence, DISC-UCN.
This project has been supported by the LP2021-9 Lend\"ulet grant of the Hungarian Academy of Sciences.
This work benefited from a workshop supported by the National Science Foundation under Grant No. OISE-1927130 (IReNA), the Kavli Institute for Cosmological Physics, and the University of Chicago Data Science Institute. 

Funding for the Sloan Digital Sky Survey V has been provided by the Alfred P. Sloan Foundation, the Heising-Simons Foundation, the National Science Foundation, and the Participating Institutions. SDSS acknowledges support and resources from the Center for High-Performance Computing at the University of Utah. SDSS telescopes are located at Apache Point Observatory, funded by the Astrophysical Research Consortium and operated by New Mexico State University, and at Las Campanas Observatory, operated by the Carnegie Institution for Science. The SDSS web site is \url{www.sdss.org}.

SDSS is managed by the Astrophysical Research Consortium for the Participating Institutions of the SDSS Collaboration, including the Carnegie Institution for Science, Chilean National Time Allocation Committee (CNTAC) ratified researchers, Caltech, the Gotham Participation Group, Harvard University, Heidelberg University, The Flatiron Institute, The Johns Hopkins University, L'Ecole polytechnique f\'{e}d\'{e}rale de Lausanne (EPFL), Leibniz-Institut f\"{u}r Astrophysik Potsdam (AIP), Max-Planck-Institut f\"{u}r Astronomie (MPIA Heidelberg), Max-Planck-Institut f\"{u}r Extraterrestrische Physik (MPE), Nanjing University, National Astronomical Observatories of China (NAOC), New Mexico State University, The Ohio State University, Pennsylvania State University, Smithsonian Astrophysical Observatory, Space Telescope Science Institute (STScI), the Stellar Astrophysics Participation Group, Universidad Nacional Aut\'{o}noma de M\'{e}xico, University of Arizona, University of Colorado Boulder, University of Illinois at Urbana-Champaign, University of Toronto, University of Utah, University of Virginia, Yale University, and Yunnan University.

This work has made use of data from the European Space Agency (ESA) mission
{\it Gaia} (\url{https://www.cosmos.esa.int/gaia}), processed by the {\it Gaia}
Data Processing and Analysis Consortium (DPAC,
\url{https://www.cosmos.esa.int/web/gaia/dpac/consortium}). Funding for the DPAC
has been provided by national institutions, in particular the institutions
participating in the {\it Gaia} Multilateral Agreement.
In Figure~\ref{fig:kinematics}, we have used a {\it Gaia} image by the Gaia Data Processing and Analysis Consortium (DPAC); A. Moitinho / A. F. Silva / M. Barros / C. Barata, University of Lisbon, Portugal; H. Savietto, Fork Research, Portugal.

This paper includes data collected by the TESS mission. Funding for the TESS mission is provided by the NASA's Science Mission Directorate.

The national facility capability for SkyMapper has been funded through ARC
LIEF grant LE130100104 from the Australian Research Council, awarded to
the University of Sydney, the Australian National University, Swinburne
University of Technology, the University of Queensland, the University
of Western Australia, the University of Melbourne, Curtin University of
Technology, Monash University and the Australian Astronomical Observatory.
SkyMapper is owned and operated by The Australian National University's
Research School of Astronomy and Astrophysics.  The survey data were
processed and provided by the SkyMapper Team at ANU.  The SkyMapper node
of the All-Sky Virtual Observatory (ASVO) is hosted at the National
Computational Infrastructure (NCI).  Development and support of the
SkyMapper node of the ASVO has been funded in part by Astronomy Australia
Limited (AAL) and the Australian Government through the Commonwealth's
Education Investment Fund (EIF) and National Collaborative Research
Infrastructure Strategy (NCRIS), particularly the National eResearch
Collaboration Tools and Resources (NeCTAR) and the Australian National
Data Service Projects (ANDS).

This publication makes use of data products from the Two Micron All
Sky Survey, which is a joint project of the University of Massachusetts
and the Infrared Processing and Analysis Center/California Institute of
Technology, funded by the National Aeronautics and Space Administration
and the National Science Foundation.

This publication makes use of data products from the Wide-field Infrared
Survey Explorer, which is a joint project of the University of California,
Los Angeles, and the Jet Propulsion Laboratory/California Institute of
Technology, funded by the National Aeronautics and Space Administration.

This research has made use of the VizieR catalog access tool, CDS,
Strasbourg, France.  The original description of the VizieR service was
published in Ref. \cite{och00}.

This research has made use of NASA’s Astrophysics Data System Bibliographic Services; the arXiv preprint server operated by Cornell University; and the SIMBAD databases hosted by the Strasbourg Astronomical Data Center.

\item[Author Contributions] 
APJ led the conceptualization, MIKE observations, stellar parameter and chemical abundance analysis, Population III analysis, writing, and interpretation. VC led the BOSS and kinematic analysis and contributed to writing and interpretation. SMT led the NLTE abundance analysis. ZZ led the literature compilation, total metallicity calculations, and contributed to the Population III analysis and interpretation. SMT and ZZ contributed to the BOSS target selection. PE computed the 3D models and led the 3D LTE carbon analysis. KCS led the distance determinations and contributed to stellar parameters, writing, and interpretation. HDA, HD, NMO, RT, and PNT contributed to the MIKE observations and the stellar parameter, chemical abundance, kinematic analysis, and interpretation. KS contributed to the stellar parameter analysis, writing, and interpretation. MH led the asteroseismology analysis. JT contributed to the carbon evolutionary corrections. MB contributed to the 3D and NLTE analyses. ARC and JAJ contributed to the stellar parameter analysis. All authors contributed to the manuscript, interpretation, SDSS-V infrastructure, and/or the SDSS-V high-resolution followup program.

\item[Author Information] Correspondence and requests for materials should be addressed to APJ (alexji@uchicago.edu).

\item[Data Availability] 
The BOSS spectrum of \umpstar (\texttt{sdss\_id} 95803549) will become publicly available in SDSS Data Release 20, as will the background halo star sample of \code{MINESweeper} results.
The individual line measurements, normalized MIKE spectrum, and literature star abundances and kinematics are available at DOI 10.5281/zenodo.18483957\cite{ZenodoLink}.

\item[Code Availability] 
Most codes used for analysis are publicly available on github, including
{\code{LESSPayne}}\cite{Ji2025ascl}, 
{\code{MOOG}}\cite{Sneden1973,Sobeck2011,Sneden2012}, 
{\code{TSFitPy}}\cite{Gerber2023},
{\code{Turbospectrum}}\cite{Plez2012}, and
{\code{agama}}\cite{Vasiliev2019}.
The exception is that {\code{M3DIS}}\cite{Eitner2024} and the version of \code{Multi3D} used are not public yet, but a public release is planned.

\item[Competing Interests] The authors declare no competing interests.

\end{addendum}

\bibliography{masterbiblio}

\let\thefootnote\relax\footnote{

\begin{affiliations}
\item Department of Astronomy \& Astrophysics, University of Chicago, 5640 S. Ellis Avenue, Chicago, IL 60637, USA
\item Kavli Institute for Cosmological Physics, University of Chicago, 5640 S Ellis Avenue, Chicago, IL 60637, USA
\item NSF-Simons AI Institute for the Sky (SkAI), 172 E. Chestnut St., Chicago, IL 60611, USA
\item Center for Astrophysics $|$ Harvard \& Smithsonian, 60 Garden St, Cambridge, MA 02138, USA
\item Max Planck Institute for Astronomy, 69117 Heidelberg, Germany
\item Heidelberg University, Grabengasse 1, 69117 Heidelberg, Germany
\item William H. Miller III Department of Physics \& Astronomy, Johns Hopkins University, 3400 N Charles St, Baltimore, MD 21218, USA
\item Department of Physics and Astronomy, Vanderbilt University, 6301 Stevenson Center Ln., Nashville, TN 37235, USA
\item Department of Astronomy and Center for Cosmology and AstroParticle Physics, The Ohio State University, Columbus, OH 43210, USA
\item Department of Astronomy, University of Florida, Bryant Space Science Center, Stadium Road, Gainesville, FL 32611, USA
\item School of Physics \& Astronomy, Monash University, Wellington Road, Clayton, Victoria 3800, Australia
\item Center for Computational Astrophysics, Flatiron Institute, 162 5th Avenue, New York, NY 10010, USA
\item Space Telescope Science Institute, 3700 San Martin Drive, Baltimore, MD 21218, USA
\item Department of Astronomy, Yale University, New Haven, CT 06520, USA
\item Universidad Cat\'olica del Norte, N\'ucleo UCN en Arqueolog\'ia Gal\'actica - Inst. de Astronom\'ia, Av. Angamos 0610, Antofagasta, Chile
\item Universidad Cat\'olica del Norte, Departamento de Ingenier\'ia de Sistemas y Computaci\'on, Av. Angamos 0610, Antofagasta, Chile
\item Department of Astronomy, The University of Texas at Austin, 2515 Speedway Boulevard, Austin, TX 78712, USA
\item The Observatories of the Carnegie Institution for Science, 813 Santa Barbara Street, Pasadena, CA 91101, USA
\item LIRA, Observatoire de Paris, Universit\'e PSL, Sorbonne Universit\'e, Universit\'e Paris Cit\'e, CY Cergy Paris Universit\'e, CNRS, 92190 Meudon, France
\item Institut de Ci\`{e}ncies del Cosmos (ICCUB), Universitat de Barcelona, Mart\'{i} i Franqu\`{e}s 1, E-08028 Barcelona, Spain
\item Institut d'Estudis Espacials de Catalunya (IEEC), E-08034 Barcelona, Spain
\item Kavli IPMU (WPI), UTIAS, The University of Tokyo, Kashiwa, Chiba 277-8583, Japan
\item ELTE E\"otv\"os Lor\'and University, Gothard Astrophysical Observatory, 9700 Szombathely, Szent Imre H. st. 112, Hungary
\item MTA-ELTE Lend{\"u}let ``Momentum" Milky Way Research Group, 9700 Szombathely, Szent Imre H. st. 112, Hungary
\item HUN-REN CSFK, Konkoly Observatory, Konkoly Thege Mikl\'os \'ut 15-17, Budapest, 1121, Hungary; E\"otv\"os Lor\'and University, Institute of Physics and Astronomy, P\'azm\'any P\'eter s\'et\'any 1, Budapest, Hungary
\item Department of Astronomy, University of Illinois at Urbana-Champaign, Urbana, IL 61801, USA
\item Department of Physics, Montana State University, P.O. Box 173840, Bozeman, MT 59717, USA
\item Center for Astrophysics and Space Astronomy, University of Colorado, 389 UCB, Boulder, CO 80309-0389, USA
\item Department of Astronomy \& Astrophysics, The Pennsylvania State University, University Park, PA 16802, USA
\item The Institute for Gravitation for and the Cosmos, The Pennsylvania State University, University Park, PA 16802, USA
\end{affiliations}
}

\end{document}